\documentclass[preprint,a4paper]{aastex}

\usepackage{graphics,epsf}
\usepackage{amsmath}                
\usepackage{amsfonts}               
\usepackage{amssymb}                
\usepackage{epsfig}
\usepackage{rotating}

\def \cm{~\rm{cm}}
\def \s{~\rm{s}}
\def \km{~\rm{km}}

\def \K{~\rm{K}}
\def \g{~\rm{g}}

\def \AU{~\rm{AU}}

\def \yr{~\rm{yr}}

\def \days{~\rm{day}}

\def \keV{~\rm{keV}}
\def \astrobj#1{#1}

\begin{document}


\title{Possible Implications of Mass Accretion in Eta Carinae}

\author{Amit Kashi\altaffilmark{1} and Noam Soker\altaffilmark{1}}

\altaffiltext{1}{Department of Physics, Technion$-$Israel
Institute of Technology, Haifa 32000 Israel;
kashia@physics.technion.ac.il;
soker@physics.technion.ac.il.}

\begin{abstract}

We apply the previously suggested accretion model for the behavior of
the super-massive binary system \astrobj{$\eta$ Car} close to
periastron passages. In that model it is assumed that for $\sim
10$~weeks near periastron passages one star is accreting mass from the
slow dense wind blown by the other star. We find that the secondary,
the less massive star, accretes $\sim 2 \times 10^{-6} M_{\odot}$. This
mass possesses enough angular momentum to form a disk, or a belt, around
the secondary. The viscous time is too long for the establishment of
equilibrium, and the belt must be dissipated as its mass is being blown
in the reestablished secondary wind. This processes requires about half
a year, which we identify with the recovery phase of \astrobj{$\eta$
Car}. We show that radiation pressure, termed radiative braking, cannot
prevent accretion. In addition to using the commonly assumed binary
model for \astrobj{$\eta$ Car}, we also examine alternative models
where the stellar masses are larger, and/or the less massive secondary
blows the slow dense wind, while the primary blows the tenuous fast
wind and accretes mass for $\sim 10~$week near periastron passages. We
end by some predictions for the next event (January-March 2009).

\end{abstract}



\section{INTRODUCTION}
\label{sec:intro}

\astrobj{$\eta$ Car} is a massive binary system with some basic
undetermined parameters and open questions. Its $5.54 \yr$ periodicity
is observed from the radio (Duncan \& White 2003), through the IR
(Whitelock et al. 2004), visible (e.g., van Genderen et al. 2006) and
to the X-ray band (Corcoran 2005). The present period was determined
recently to be $P_{\rm pres} =2022.7 \pm 1.3~$d (Damineli et al.
2008a). According to most models, the periodicity follows the
5.54~years periodic change in the orbital separation in this highly
eccentric, $e \simeq 0.9-0.95$, binary system (e.g., Hillier et al.
2006). The X-ray cycle presents a deep minimum lasting $\sim 70$~day
(Corcoran 2005 and references therein) and occurring more or less
simultaneously with the spectroscopic event. The spectroscopic event is
defined by the fading, or even disappearance, of high-ionization
emission lines (e.g., Damineli 1996; Damineli et al.\ 1998, 2000,
2008a,b; Zanella et al. 1984). The spectroscopic event includes changes
in the continuum and lines (e.g., Martin et al. 2006,a,b; Davidson et
al. 2005; Nielsen et al. 2007; van Genderen et al. 2006; Damineli et
al. 2008b). The X-ray minima and the spectroscopic events are assumed
to occur near periastron passages.

It is generally agreed that the orbital plane lies in the equatorial
plane of the bipolar structure$-$the Homunculus (Davidson et al. 2001).
However, there is a disagreement on the orientation of the seimajor axis in
orbital plane, with some groups arguing that the secondary (less massive) star
is away from us during periastron passages (e.g., Nielsen et al. 2007;
Damineli et al. 2008b), while others argue the secondary is toward us during periastron
passages (Falceta-Goncalves et al. 2005; Abrham et al. 2005; Kashi \& Soker 2007b).
Other semimajor axis orientations have also been proposed
(Smith et al. 2004; Dorland 2007).
In any case, our present research does not depend on the orientation.

The usual assumption, the `common model', is that the more massive and
more luminous star is the cooler one and the one that blows the dense
slow wind (see section \ref{sec:profile} below). We will also use the
common model as our basic model. However, as our study is more general,
and due to many uncertainties in the properties of \astrobj{$\eta$
Car}, we will consider other possibilities, applicable to
\astrobj{$\eta$ Car} (Kashi et al. 2008), and to other more general
cases. We will consider a speculative case where the slow dense wind is
blown by the less massive star, and a case where the binary system is
more massive. Motivations for these models are discussed in Kashi et
al. (2008).

Another riddle concerns the exact chain of events that cause the spectroscopic event.
In general there is a slow variation along the entire orbit, the `slow variation'
component (Damineli et al. 2008b), and fast variations near the event, the
`collapse' component (Damineli et al. 2008b).
The slow variation is probably caused by the motion of the secondary around
the primary, getting in and out of the dense part of the primary wind.
Another effect is the slow variation of the X-ray emission.
The emitting gas is the post-shock secondary wind (Pittard \& Corcoran 2002;
Akashi et al. 2006), which becomes denser as the two stars approach each other.
While some lines and bands can be explained by the motion of the secondary around
the primary, e.g. the radio light curve (Kashi \& Soker 2007a), other require
an additional effect to account for the collapse component.
Such is the case for the $\sim 10$~week long X-ray minimum.

In a series of papers (Soker 2005, 2007a; Akashi et al. 2006; Soker \&
Behar 2006; Kashi \& Soker 2008) we proposed that the fast variations
during the spectroscopic event are caused by the collapse of the winds
interaction region onto the secondary star. The secondary then accretes
mass for $\sim 10~$weeks, and its fast wind is blown only along the
polar direction. One of the results is that the secondary wind is very
weak, explaining the fast variation and minimum of the X-ray emission
(Akashi et al. 2006). The accretion phase can explain also the fast
changes of some spectral lines (Soker 2007a), and the IR emission
(Kashi \& Soker 2008).

In the present paper we go beyond previous papers in trying to explore
the nature of the accretion phase, and its influence on the secondary
for about half a year after the event.
In \S\ref{sec:timescales} we discuss the relevant timescales for the
accretion process. In \S\ref{sec:profile} we define the system
parameters. \S\ref{sec:accmass} gives a detailed  calculation for
the accreted mass and angular momentum onto the secondary during the
periastron passage. \S\ref{sec:radiative} discuss the importance
of radiative braking in the accretion process. In
\S\ref{sec:dissipation} we study processes that can account for
the 6 months recovery pjase after the spectroscopic event. In
\S\ref{sec:alternative} we present speculative alternative binary
models and repeat the accretion calculations for these models. We
summarize in \S\ref{sec:summary}.

\section{TIME SCALES}
\label{sec:timescales}

Before considering the accretion process itself, we derive several time scales relevant
to the studied processes.

While the orbital period is $P=5.54$ years, the duration of the strong
interaction near periastron passage is much shorter. We define the
typical time scale of the strong interaction phase occurring near
periastron passage as the time the secondary crosses a distance of the
primary diameter centered at periastron. With $v_{{\rm orb}2,{\rm per}}$
being the relative velocity of the two stars at periastron this
time is defined as
\begin{equation}
t_{\rm{pp}}=\frac{2R_1}{v_{{\rm orb}2,{\rm per}}} \simeq 6
\left(\frac{R_1}{180R_{\odot}}\right) \left(\frac{v_{2,{\rm
per}}}{470\rm{km s^{-1}}}\right)^{-1} \days. \label{Tbelt}
\end{equation}

Under our assumptions the secondary accretes mass during periastron passages (Soker 2005).
This mass has angular momentum, and it accumulates around the secondary (see section
\ref{sec:accmass}). The Keplerian period on the secondary equator is
\begin{equation}
t_{\rm{Kep}}= 1.89 \left(\frac{R_2}{20R_{\odot}}\right)^{3/2}
\left(\frac{M_2}{30 M_{\odot}}\right)^{-1/2} \days.
\label{Tkep}
\end{equation}
The accretion phase in our model (Soker 2005, 2007a) lasts
$t_{\rm acc}\sim 70~$days, as the length of the X-ray minimum (Akashi
et al. 2006). Despite the fact that $t_{\rm{Kep}} \ll t_{\rm acc}$ a
geometrically thin accretion disk is not formed. The reason is that the
viscosity time scale in the disk $ t_{\rm visc}$ is very long. In the
$\alpha$ disk model where the viscosity is $\nu = \alpha C_s H$
(Shakura \& Sunyaev 1973), where $C_s$ is the sound speed and $H(R)$
the disk hight, the viscosity time scale is given by
\begin{equation}
\begin{split}
t_{\rm{visc}}=\frac{R^2}{\nu} = \frac{R^2}{\alpha C_s H} \simeq
t_{\rm{Kep}} \left(\frac{R}{H}\right)^{2} \frac{1}{2 \pi \alpha}  \\
=30 \left(\frac{R}{10 H}\right)^{2}
\left(\frac{R_2}{20R_{\odot}}\right)^{3/2} \left(\frac{M_2}{30
M_{\odot}}\right)^{-1/2} \frac{1}{\alpha}\days , \label{Tvisc}
\end{split}
\end{equation}
where the equation was scaled near the secondary equator, where the
viscosity time scale is the shortest. Since we expect $\alpha<1$, the
viscosity time is not short enough to make a geometrically thin disk
during the accretion time $t_{\rm {acc}}$. However, the viscosity is
not completely negligible and the accreted mass will form a thick accretion
disk attached to the secondary$-$a \emph{belt}.

To summarize this section, we see that the relation among the timescales is such that
\begin{equation}
t_{\rm Kep} \la t_{\rm{pp}} \ll t_{\rm{\rm acc}} \sim t_{\rm{visc}} \ll
P. \label{Ts}
\end{equation}
We therefore conclude that during the strong interaction phase, lasting $\sim 1~$week,
when the accretion rate is very high (see section \ref{sec:accmass}) the accretion belt
will be far from equilibrium.
During the entire accretion phase, lasting $\sim 10~$weeks, a thick accretion disk might
be formed, but not a geometrically thin one.

\section{THE BINARY SYSTEM IN THE COMMON SMALL-MASSES MODEL}
\label{sec:profile}

The \astrobj{$\eta$ Car} binary parameters used by us are based on
several papers and taking into account the present disagreement on some
of the binary parameters (e.g., Ishibashi et al. 1999; Damineli et al.
2000; Corcoran et al. 2001, 2004b; Hillier et al. 2001; Pittard \&
Corcoran 2002; Smith et al. 2004; Verner et al. 2005; Damineli et al.
2008a, b; Dorland 2007). This is termed by us the `small-masses model'.
The assumed stellar masses are $M_1=120 M_\odot$, $M_2=30 M_\odot$, the
eccentricity is $e=0.9-0.93$, and orbital period $P=2024$~days
(Damineli et al. 2008 suggested recently that $P=2022.7 \pm 1.3$~days),
hence the semi-major axis is $a=16.64 \AU$, and the orbital separation
at periastron is $r=1.66-1.16 \AU$ for $e=0.9-0.93$, respectively. The
stellar radii are taken to be $R_1=180R_\odot$ and $R_2 = 20 R_\odot$.
The mass loss rates are $\dot M_1=3 \times 10^{-4} M_\odot \yr^{-1}$
and $\dot M_2 =10^{-5} M_\odot \yr^{-1}$. The terminal wind speeds are
taken to be $v_1=500 \km \s^{-1}$ and $v_2=3000 \km \s^{-1}$. It is
assumed here that the orbital plane is oriented in the same plane as
the equatorial zone of the primary's wind, and the equatorial plane of
the Homunculus. One has to remember that all these parameters are
uncertain to some degree. The most relevant to us is the large
uncertainty in the acceleration zone, i.e., the velocity profile of the
primary wind.

Because the secondary gets to be very close to the primary near periastron passages,
the interaction of the secondary with the primary wind strongly depends on the
acceleration zone of the primary wind.
The primary's wind velocity profile can be described using the $\beta$-profile:
\begin{equation}
v_1(r)=v_0+(v_{\infty}-v_0)\left(1-\frac{R_1}{r}\right)^{\beta} ,
\label{v1}
\end{equation}
where $v_0=20 \km \s^{-1}$ is the sound velocity on the primary's
surface, $v_{\infty}=500 \km \s^{-1}$ is the primary's wind terminal
velocity, and $\beta$ is a parameter of the wind model.
A recent study by Kraus et al. (2007) suggest that in OB-supergiants the
range can be $\beta \simeq 1-3$.
We will consider the two extreme values of $\beta=1$ and $\beta=3$.

\section{THE ACCRETION DURING PERIASTRON PASSAGE}
\label{sec:accmass}
\subsection{Mass accretion rate}
\label{sec:mass}

Benz \& Hills (1992) performed Smooth-Particle Hydrodynamics (SPH)
simulations of main-sequence stars whose masses differ by a factor of
$5$. In one of their grazing encounter simulations the lower-mass star
accreted an atmosphere of gas that formed a ring around it. A `tongue'
of gas connecting the two stars was also present at that time. In
another study, Yamada et al. (2008) performed high resolution SPH
simulations of tidal encounter of a red giant star with a main sequence
star. Their simulations showed that for some parameters of the
interacting binary system the main sequence star can accrete $\sim
10^{-3} M_{\odot}$ from the envelope of the red giant star, mainly as a
result of tidal interaction. The red giant in the model of Yamada et
al. (2008) does not blow any stellar wind; a stellar wind could further
enhance the accreted mass. The two studies described here motivate us
to consider the accretion of mass with high specific angular momentum
in each periastron passage in \astrobj{$\eta$ Car}, as there are many
similarities between the seemingly different processes of stars
collision described above and of \astrobj{$\eta$ Car}.

At periastron the orbital separation between the two stars in
\astrobj{$\eta$ Car} is $r \sim 1.6 \AU$ (depending on the eccentricity
$e$, and masses). This is so close that the secondary actually passes
through the dense acceleration zone of the primary's wind. This
geometry is quite similar to stellar collision, although the
`colliding' stars in \astrobj{$\eta$ Car} are very different in mass
and physical properties from the colliding main-sequence stars studied
by Benz \& Hills (1992).

In proposing the accretion model for the spectroscopic event (Soker
2005) the accretion rate was simply estimated using the
Bondi-Hoyle-Lyttleton (BHL) accretion prescription. This accretion
process is for a point mass moving through a more or less homogeneous
medium. However, here we have the secondary moving inside the
acceleration zone of the primary wind, and a mass transfer process
similar to a Roche lobe overflow (RLOF) might occur for several days.
(e.g., Dorland 2007). It will not be a full RLOF because the primary
spin and orbital motion are not synchronized near periastron passage.
Therefore, very close to periastron passage, $\vert t \vert \la 10~$day,
the accretion process will be an hybrid of the BHL and the
RLOF mass transfer processes.
At the end of the accretion phase the accretion will be of the BHL type
see figure \ref{BHL} and \ref{RL}.
Because of the very complicated geometry of the colliding winds,
we can only estimate the mass and angular momentum accretion rates.
To be able to make these estimates, we make the following steps.

The radial (along the line joining the two stars) component of the
relative velocity between the secondary star and the primary's wind is
$v_1-v_r$, where $v_r$ the radial component of the orbital velocity;
$v_r$ is negative when the two stars approach each other. The total
relative speed between the secondary and the primary's wind is
\begin{equation}
v_{\rm wind1} = \left[v_\theta^2 + (v_1-v_r)^2 \right]^{1/2},
\label{vwind1}
\end{equation}
where $v_\theta$ is the tangential component of the orbital velocity
(see figure 1 of Soker 2005). We neglect any time delay between
ejection of the wind by the primary and its accretion onto the
secondary. This assumption is justified over most of the orbit, but not
near periastron when the secondary is inside the acceleration zone of
the primary wind and the relative orbital velocity is not much smaller
than the primary wind velocity, or even larger. Nevertheless, we make
this assumption as the uncertainty it introduces is much smaller than
the other uncertainties, and its influence on the results is small. In
particular, the tide raised by the secondary on the primary surface
(Dorland 2007) is likely to affect the mass loss process in causing a
departure from spherical symmetry and in enhancing the mass loss rate.
In other words, we don't modify the primary surface according to the
gravitational potential of the secondary (which when synchronization
exists is the Roche potential). A clear example for such tidal
interaction was presented by Yamada et al. (2008). We don't consider
this process here because it is much too complicated. Its effect will
be to increase the mass accretion rate by the secondary, hence making
the processes studied by us more significant.

The secondary's orbit is parameterized by the orbital angle $-\pi <
\theta \le \pi$, with $\theta=0$ at periastron. The center of the
secondary is at a distance
\begin{equation}
r(\theta)=\frac{a(1-e^2)}{1+e \cos \theta} \label{rtheta}
\end{equation}
from the center of the primary, where $a$ is the semi-major axis of the
elliptic orbit and $e=0.9-0.93$ is the eccentricity.

As discussed in section \ref{sec:intro} here (Soker 2005; Akashi et al.
2006) during the accretion phase the secondary accretes from the
primary wind. The effective accretion radius of the secondary $R_{\rm
acc}$, i.e., the radius from where mass is accreted, depends on several
parameters, in particular on the orbital separation $r (\theta)$. We
consider the two limits of RLOF and BHL accretion. We don't really
consider the flow of gas in a RLOF type situation. We rather take the
circular accretion radius to be equal to the Roche lobe equivalent radius,
(Eggleton 1983) to give some weight to the gravity of the two stars.
This radius is given by
\begin{equation}
R_{RL}(\theta)=\frac{0.49q^{2/3}}{0.6q^{2/3}+\ln(1+q^{1/3})}r(\theta)=0.2676r(\theta)
\label{roche}
\end{equation}
where $q=M_2/M_1=0.25$ is the mass ratio in the common model. We note
again that we use this approximation for the RL radius even that the
system is not synchronized.

In the BHL accretion from the wind the effective accretion radius is
according to Bondi \& Hoyle (1944) when the flow is highly supersonic,
as is the case here,
\begin{equation}
R_{BHL}(\theta)=\frac{2GM_2}{v_{\rm wind1}^2} \label{bondi}.
\end{equation}

\begin{figure}[!ht]
   \centering
   \parbox{\textwidth}{
      \hspace{+0.0cm}\includegraphics[width=0.49\textwidth]{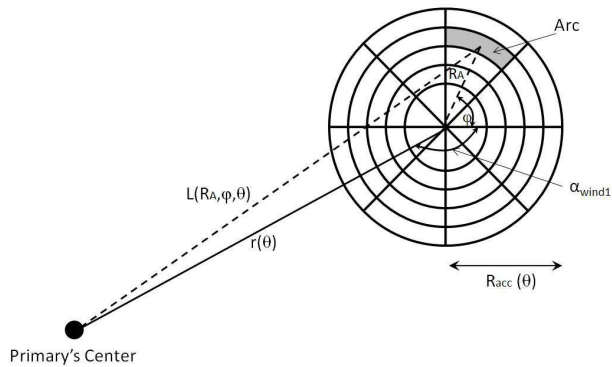}}
      \caption{\footnotesize The profile of the accretion cross section and some quantities defined in the text.
      Near periastron passage, i.e., during the strong interaction phase,
      the secondary moves more or less perpendicular to the plane of the page.}
      \label{numprofile}
\end{figure}

\begin{figure}[!ht]
   \centering
   \parbox{\textwidth}{
      \hspace{+0.0cm}\includegraphics[width=0.49\textwidth]{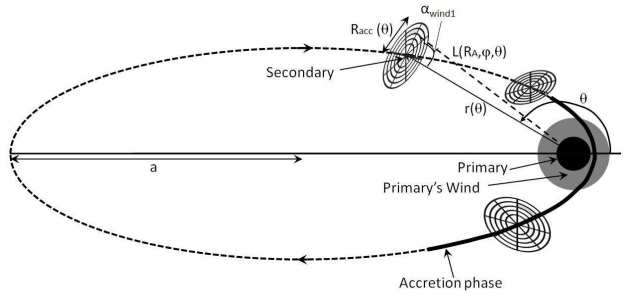}}
      \caption{\footnotesize Schematic drawing of the geometry of the orbit and the accretion
      cross section used in the calculations. The gray donut around the primary
       represents the radius where the primary's wind accelerated to $100 \km \s^{-1}$, for wind
       profile with $\beta=3$.
       The orbit and binary star are drawn for scale for the case $e=0.93$ and $R_1=180 R_\odot$. }
      \label{numgeometry}
\end{figure}

\begin{figure}[!ht]
\resizebox{0.39\textwidth}{!}{\includegraphics{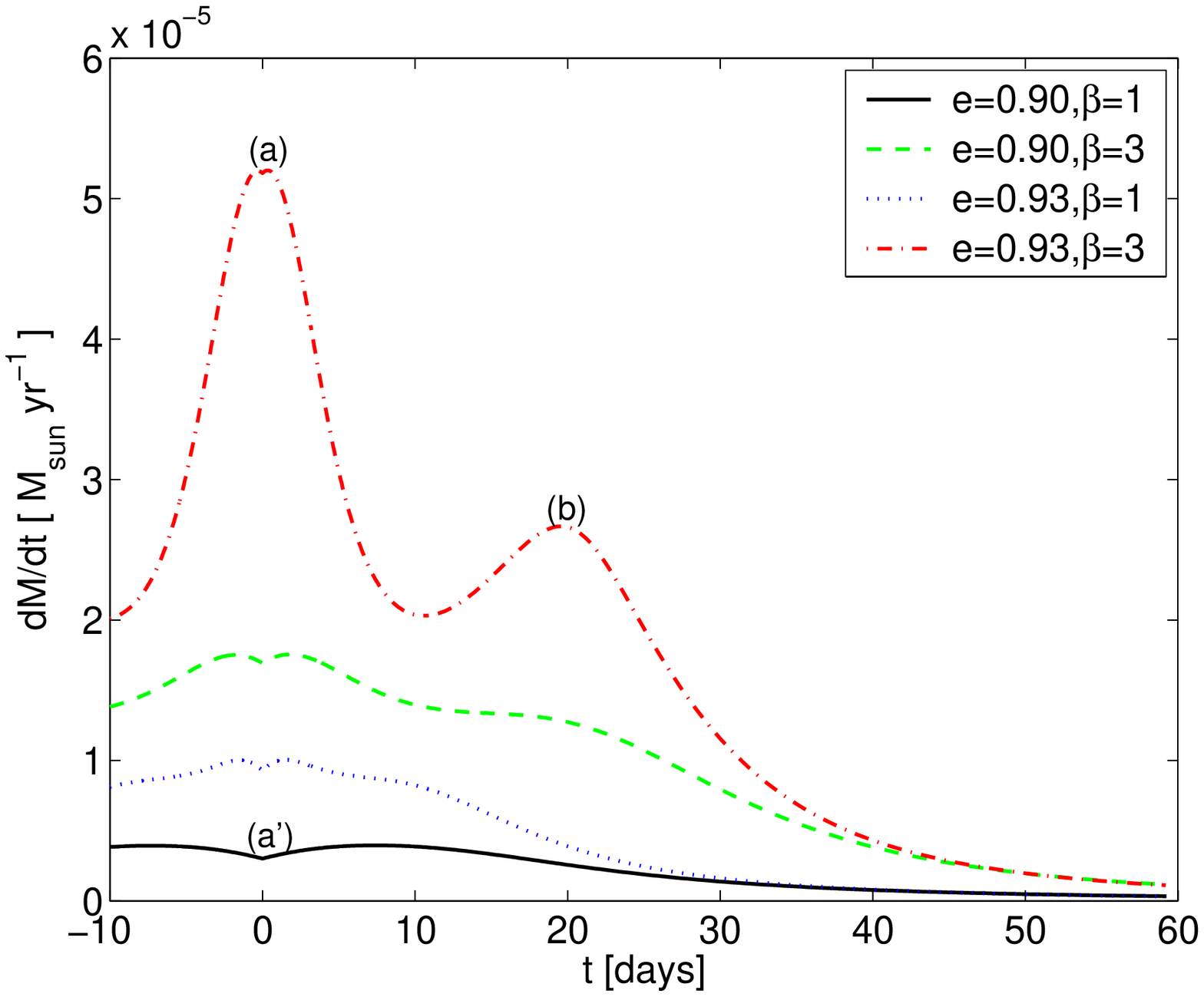}}
\resizebox{0.39\textwidth}{!}{\includegraphics{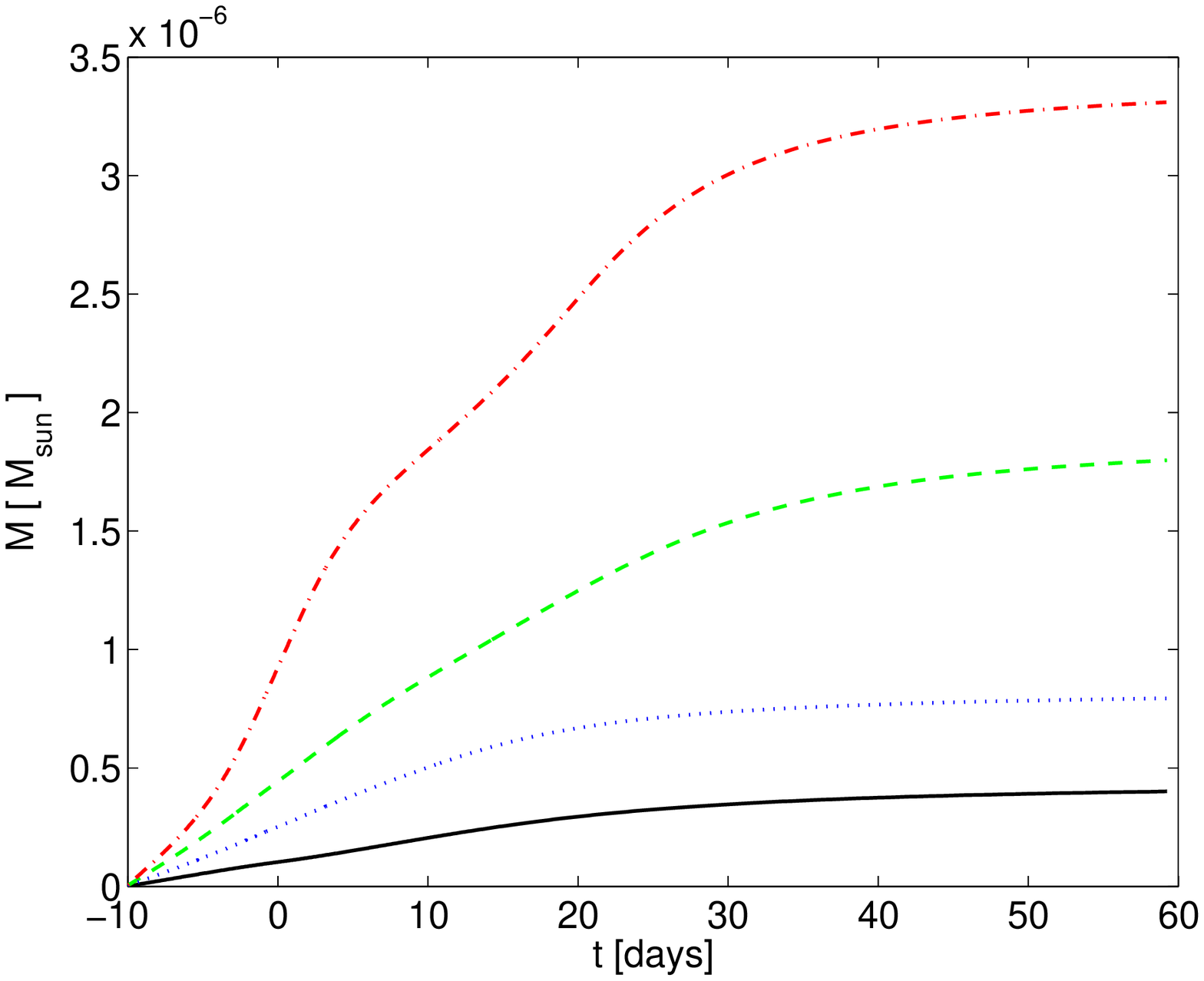}}
\caption{\footnotesize The accretion rate (first panel) and accreted
mass (second panel) as a function of the time near periastron passage
obtained using Bondi-Hoyle-Lyttleton accretion radius (first four rows
in Table \ref{Table:Macc}).
At close separation the wind's density is high. This accounts for the
wide maximum in mass accretion rate around $t=0$.
Very close to periastron the relative velocity between the wind and the
accreting star is large. As a result of that the accretion radius is
small, and mass accretion low. This accounts for the local minimum at $t=0$
in the three lower plots (point $a^\prime$).
For the case of very slow wind and close periastron $(e,\beta)=(0.93,3)$,
the higher density effect dominates at all times, and there is no local
minimum at $t=0$ (point $a$).
As the two stars recede each other the relative velocity between the wind
and accreting star reach a minimum, and therefore the accretion radius
reaches a maximum (see equation \ref{bondi}). This explains the maximum at point $b$.
}
\label{BHL}
\end{figure}

\begin{figure}[!ht]
\resizebox{0.39\textwidth}{!}{\includegraphics{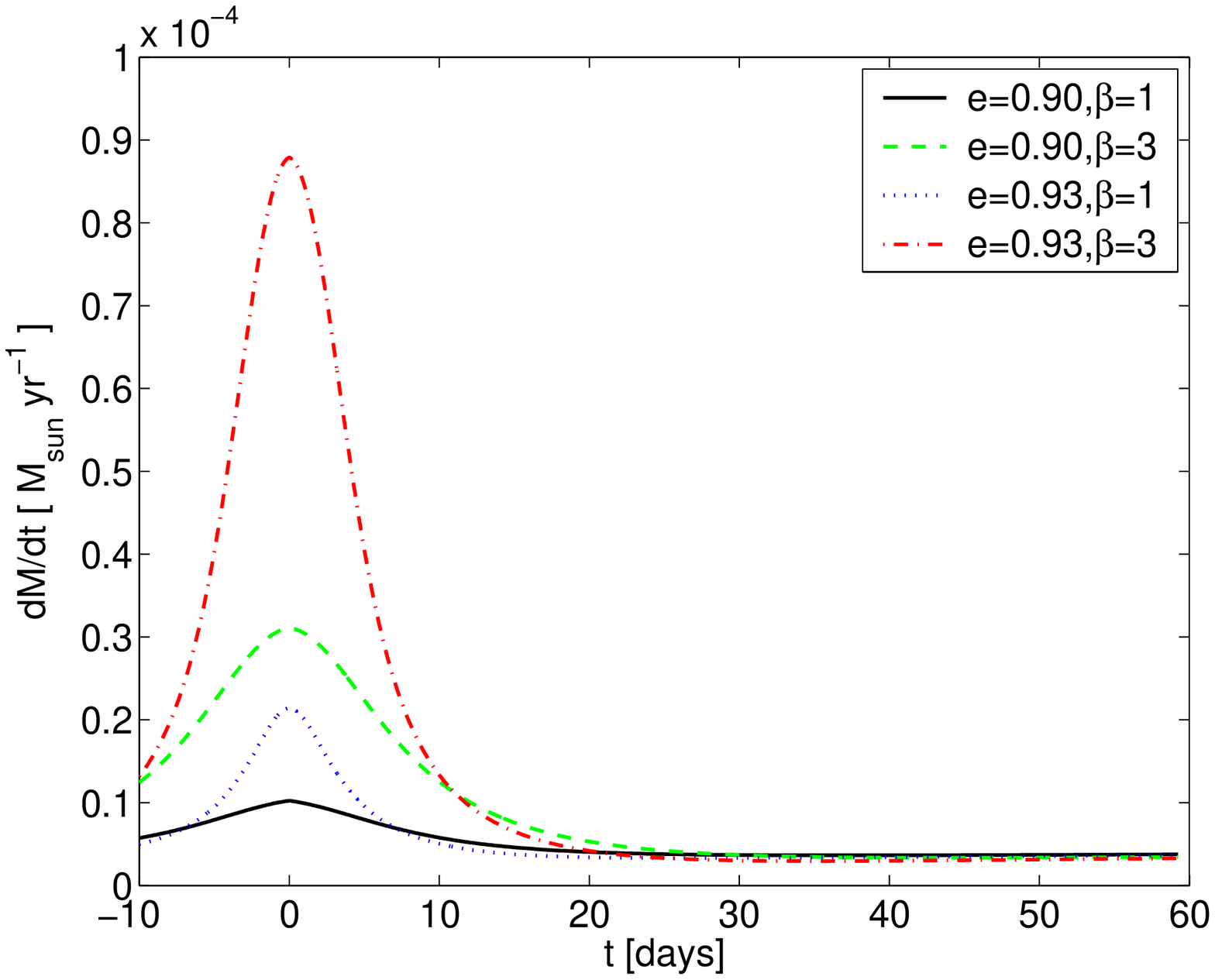}}
\resizebox{0.39\textwidth}{!}{\includegraphics{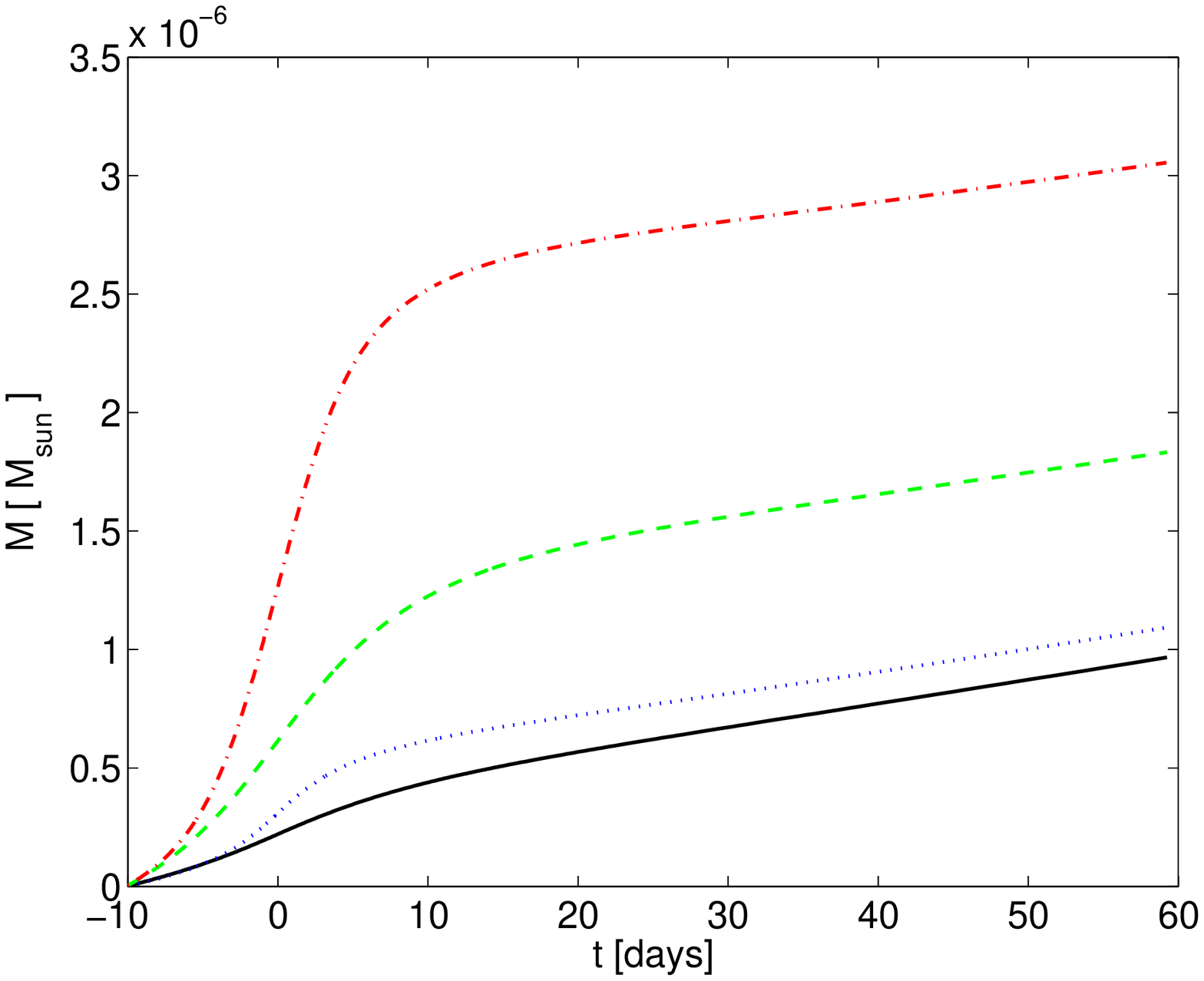}}
\caption{\footnotesize The accretion rate (first panel) and accreted
mass (second panel) as a function of the time near periastron passage
obtained using secondary Roche lobe (RL) as accretion radius (last four rows
in Table \ref{Table:Macc}). The maximum accretion rate at $t=0$ is
explained by a maximum of the density $\rho$.
The usage of the RL radius is applicable only very close to periastron,
$\vert t \vert \la 10$~day. At the end of the accretion phase the BHL
accretion must be used. }
\label{RL}
\end{figure}

We define the accretion plane to be perpendicular to the orbital plane
and rotated in an angle $\alpha_{\rm{wind1}}$ from the line connecting
the two stars, away from the primary, where
\begin{equation}
\tan(\alpha_{\rm{wind1}})=\frac{v_\theta}{v_1-v_r} \label{alphawind1}.
\end{equation}
We define the angle $\varphi$ as the azimuthal angle in that plane,
measured from the direction away from the primary, as drawn
schematically in Figure \ref{numprofile}. In both cases (BHL and RLOF)
near periastron $R_{\rm acc}$ is not much smaller than the orbital
separation. Therefore, the density of the primary wind cannot be
treated as a constant in the cross section of the accretion
cylinder$-$the circle of radius $R_{\rm acc}$ drawn in Figure
\ref{numprofile}. To calculate the dependance of the primary wind
density on $\varphi$ and on the distance from the secondary, we slice
the cross section to $30 \times30$ arcs, as drawn in Figure
\ref{numprofile}. The accreted mass from each arc is calculated
separately according to the density at that point, and it is added to
the accreted mass. Each arc has coordinates $(R_{\rm
A},\varphi,\theta)$: $(R_{\rm A},\varphi)$ are the coordinates relative
to the secondary, where $R_{\rm A}$ is the distance from the secondary,
and $\theta$ is the location of the secondary in the orbit. These
quantities are defined also in Figure \ref{numgeometry}.

The distance of an arc from the primary is given by
\begin{equation}
L(R_{\rm A},\varphi,\theta)^2=r^2(\theta)+R^2_{\bf A} - 2 r(\theta)
R_{\rm A} \cos \varphi \cos(\alpha_{\rm{wind1}}). \label{Lr1}
\end{equation}
As before, we neglect the delay between the position of the primary
when it ejected the wind and its present location, such that the
primary wind density is given
\begin{equation}
\rho(R_{\rm A},\varphi,\theta)=\frac{\dot M_1}{4 \pi L^2 v_1}
\label{rho}.
\end{equation}

The area of each arc is given by $A(R_{\rm
A},\varphi,\theta)=\frac{d\varphi}{2} (R_{\rm A}^2-R_{\rm A,in}^2)$
where in our calculation $d\varphi=2\pi/30$ and $R_{\rm A,in}=R_{\rm
A}-R_{acc}/30$. As the secondary moves on its orbit from $\theta$ to
$\theta+d\theta$ the arc moves  an azimuthal distance $v_{\rm wind1}
dt$ where $dt$ is the corresponding time it takes for the secondary to
pass that orbital angle.

The volume swept by the arc in each evolutionary step is $dV =A v_{\rm
wind1} dt$ The mass accreted by each arc is therefore $dM_A =dV
\rho(L)$. The total accreted mass as a function of the orbital angle is
\begin{equation}
dM(\theta)=\sum_{R_{\rm A}}\sum_{\varphi}dM_A.
\end{equation}
 {} From this quantity we can derive the mass accretion rate by
$\dot{M}_{\rm acc}=dM(\theta)/dt(\theta)$, where $dt$ is the time it
takes the secondary to move the orbital angle $d \theta$. The total
accreted mass along the orbit is
\begin{equation}
M(\theta)=\sum dM(\theta) ,
\end{equation}
where summation is from $\theta(-10 \days)$ to $\theta(+60 \days)$,
which is the time interval when accretion is assumed to take place
(Akashi et al. 2006). The maximum of $M(\theta)$ is the total accreted
mass $M_{\rm{acc}}$.

The calculation was processed for the eight cases summarized in Table
\ref{Table:Macc}. The variation of the accretion rate and accreted mass
with time are presented in Figure \ref{BHL} for the BHL type accretion
flow, and in Figure \ref{RL} for the case where the accretion radius is
taken from the Roche lobe geometry to mimic RLOF type accretion. We
emphasize again that what we term RLOF is not a real RLOF. It is rather
estimation of the accretion rate by taking the accretion radius to be
equal to the average radius of the secondary's Roche lobe.
\begin{table}[!ht]

\begin{tabular}{||c||c|c|c|c|c||}
\hline \hline
Case&Accretion&$e$&$\beta$&$M_{\rm{acc}}$&$\underline{j_{\rm t-acc}}$\\
 &Mode& & &$(10^{-6} M_{\odot})$&${j_2}$\\
\hline \hline

1&BHL&$0.9$&$1$&$0.40$&$0.41$\\

2&BHL&$0.9$&$3$&$1.80$&$0.98$\\

3&BHL&$0.93$&$1$&$0.79$&$0.81$\\

4&BHL&$0.93$&$3$&$3.31$&$1.29$\\

5&RL&$0.9$&$1$&$0.97$&$0.87$\\

6&RL&$0.9$&$3$&$1.83$&$0.93$\\

7&RL&$0.93$&$1$&$1.09$&$0.93$\\

8&RL&$0.93$&$3$&$3.05$&$0.91$\\

\hline \hline
\end{tabular}
\caption{\footnotesize \rm{The accreted mass and specific angular
momentum for the studied cases. In the second column the accretion mode
according to which the accretion radius is calculated: Roche lobe
overflow (RL) or Bondi-Holye-Lyttleton (BHL). The third column is the
eccentricity of the binary system. In the fourth column the value of
$\beta$ in the primary's wind velocity profile is given (equation
\ref{v1}). In the fifth and sixth columns the results of the
calculations are given: the total accreted mass during the assumed 10
weeks accretion phase, and the specific angular momentum of this mass
$j_{\rm t-acc}$, respectively. $j_{\rm t-acc}$ is given in units of the
specific angular momentum of a circular Keplerian orbit on the
secondary equator, $j_2$. }} \label{Table:Macc}
\end{table}

As can be seen in Table \ref{Table:Macc} the accreted mass for $e=0.93$
is larger than for $e=0.90$. This is expected as the secondary `dives'
deeper into the primary dense wind zones as the eccentricity increases.
Over all, the accreted mass is $M_{\rm acc} \sim 0.4 - 3.3 \times10^{-6}
M_\odot$ for the cases studied here. Our calculations might
underestimate somewhat the accretion rate, mainly because we neglect
tidal effects that are expected to be significant (van Genderen \&
Sterken 2007; Dorland 2007). On the other hand we have neglected the
ram pressure exerted by the secondary wind after accretion has started,
and the delay between the ejection of the primary wind and the
accretion (the primary changes its position during that time); both
effects will somewhat reduce the mass loss rate. Still, our
calculations underestimate the mass accretion rate. We therefore will
scale the mass accretion rate with a value a little larger than the
average of the different cases studied here.

The calculations described above were repeated for different starting
times of the accretion process, but for the same time interval of $70
\days$ dictated by the x-ray minimum. We find that if the $70 \days$
accretion phase starts $20 \days$ before periastron passage ($t=-20
\days$) the mean accreted mass is higher by $\sim 10 \%$, while if it
starts at periastron ($t=0$) it is lower by $\sim 30 \%$, than the
cases we present above (where accretion starts at $t=-10 \days$).
Considering all these, we scale the accreted mass with $M_{\rm acc}
\simeq 2 \times10^{-6} M_\odot$.

We also checked the mass accretion rate using the primary's dense wind
model according to Hillier et al. (2001; 2006). These authors used a
wind model which reaches a speed of $177 \km \s^{-1}$ at a radius of
$99.4 R_\odot$, a wind speed of $375 \km \s^{-1}$ at a radius of $881
R_\odot$, a terminal wind speed of $v_1=500 \km \s^{-1}$, and a mass
loss rate of $\dot M_1=10^{-3} M_\odot \yr^{-1}$. We fit their model to
our equation (\ref{v1}) by setting $R_1=99.4 R_\odot$, $v_0=177 \km
\s^{-1}$, and a terminal wind speed of $v_1=500 \km \s^{-1}$. Using
their result $v_1(r=881 R_\odot)=375 \km \s^{-1}$ we find that
$\beta=4.08$ fits equation (\ref{v1}). Applying the previous
calculations to this wind model based on Hillier et al. (2001; 2006),
for four cases ($e=0.9,0.93$, $\beta=1,3)$, we find the average
accreted mass to be $M_{\rm acc} \simeq 1.1 \times10^{-6} M_\odot$. We
conclude that despite the pronounced differences between the wind's
model used by us and by Hillier et al., both models yield about the
same average accretion rate.

\subsection{Angular momentum}
\label{sec:angular}

Consider a compact accreting object moving through an inhomogeneous
medium, where one side is denser and/or has a higher velocity relative
to the other side of the object. This implies that the material
entering a symmetrical circular cross section, as in Figure
\ref{numprofile}, has a net angular momentum. In the steady state BHL
accretion flow the accretion column$-$the dense column behind the
accreting body$-$bent toward the side that contains less angular
momentum (Davies \& Pringle 1980). Therefore, more mass is accreted
from the low angular momentum side, and the net accreted angular
momentum is only a fraction of $\sim 0.2$ of what would be accreted if
the accretion cross section stays symmetric (Livio et al. 1986; Ruffert
1999). However, the accretion near periastron in \astrobj{$\eta$ Car}
does not reach a steady state (Akashi et al. 2006), and it is not a
pure BHL type accretion, but rather an intermediate flow between RLOF
and BHL accretion, as we argue here. The intermediate flow between RLOF
and BHL accretion assumption holds very close to periastron (for $\vert
t \vert \la 10$~day compare Figures \ref{BHL} and \ref{RL}), but later
it is only the BHL, as at large distances we don't expect the RLOF to
occur.

For that, we take the angular momentum of the accreted mass to be the
one enters the symmetric accretion cross section as drawn in Figure
\ref{numprofile}. By that we overestimate somewhat the accreted angular
momentum.

The calculation of the accreted angular momentum is more sensitive to
the assumptions we make than the calculation of the accreted mass.
In particular, in the RLOF type accretion more
mass is accreted from the first Lagrangian point $L_1$ and its
vicinity. This type of flow is neglected by us. Therefore, this effect
causes our calculation to underestimate the accreted angular momentum.
We neglect also the radial component of the orbital motion. By this we
overestimate somewhat the accreted angular momentum, because in a pure
radial motion the accreted angular momentum is zero.

Over all, we consider the calculation below as a crude, but fair,
estimate of the accreted angular momentum.

To estimate the accreted angular momentum under our assumptions we
should multiply the mass flux into each arc considered in section
\ref{sec:mass} by its specific angular momentum component perpendicular
to the orbital plane and relative to the secondary. However, this is a
complicated task in the flow structure that we study, because the
calculated angular momentum is very sensitive to the exact value of the
speed of a mass element entering the accretion cross section. After the
stars crosses periastron, the relative velocity between the secondary
and the primary's wind changes its sense, for example. We examine two
limiting cases. If the secondary moves in a straight line with a speed
$v_{\rm wind1}$ through the medium, the specific angular momentum of a
mass element relative to the secondary is $j_{A1} \simeq - v_{\rm
wind1} (R_{A} \cos \varphi)$. The minus sign was introduced so that a
positive accreted angular momentum means that it has the same direction
as the orbital angular momentum. If the mass accretion is a full RLOF,
then $j_{A2}=-R^2_{A} \Omega$. We also calculated the specific angular
momentum for a case where the secondary is in a circular orbit with a
radius equal to the periastron distance, using the results of
Wang (1981). This value is termed $j_{A3}$, and it will be given later.

In any case, considering these limiting cases, we take an expression
that gives lower accreted angular momentum than each one of them.
Namely, again we take a conservative approach to demonstrate the
importance of angular momentum. The specific angular momentum of the
gas accreted from each arc is taken to be
\begin{equation}
j_A \simeq -(R_{A} \cos \varphi)^2 \Omega, \label{ja1}.
\end{equation}
Since most mass is accreted very close to periastron, $j_{A2}$ and
$j_{A3}$ might be the most accurate out of the different expression
mentioned here for $j_A$. In the future 3D numerical simulations will
have to examine the accreted angular momentum.

\begin{figure}[!ht]
\resizebox{0.39\textwidth}{!}{\includegraphics{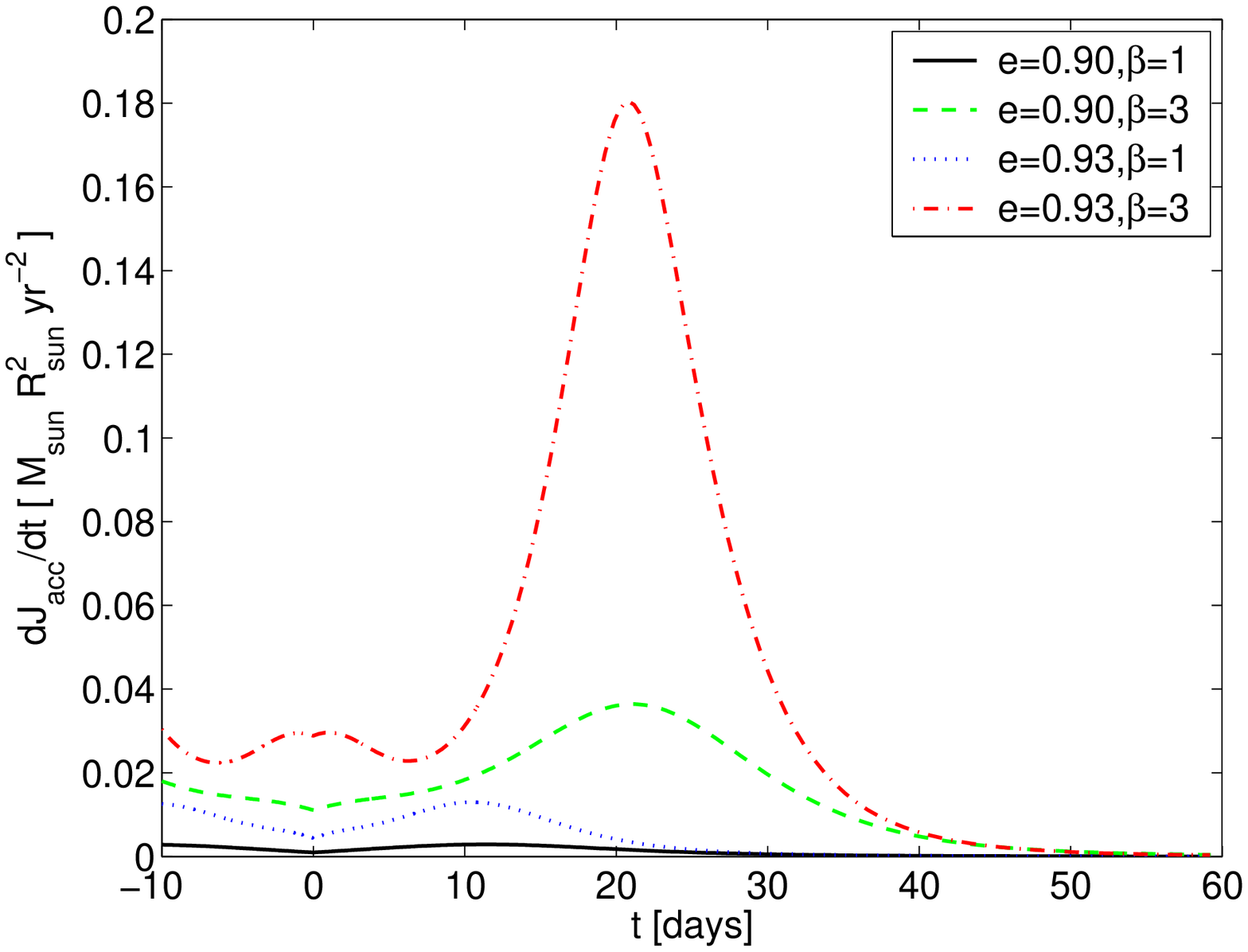}}
\resizebox{0.39\textwidth}{!}{\includegraphics{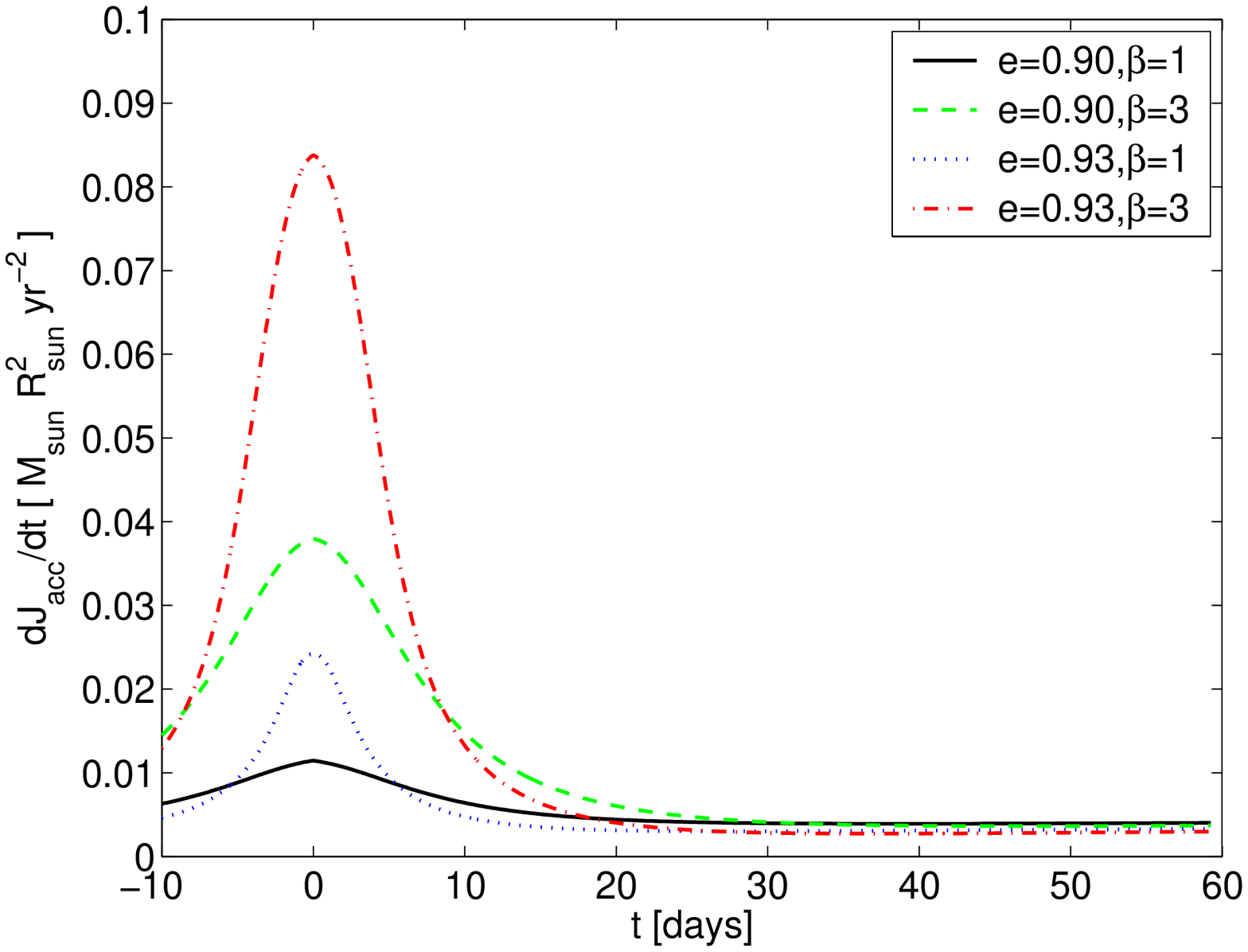}}
\caption{\footnotesize The angular momentum accretion rate as function of time.
The accretion radius is calculated according to the BHL case (upper panel),
or according to the Roche lobe radius (RLOF; second panel).
The meaning of the different lines is like in Figure \ref{BHL}.
The angular momentum accretion rate increases with increasing accretion
radius and density gradient. Approximately, the specific angular momentum depends
on accretion radius square, and linearly on the density gradient (Wang 1981).
The variations in these two quantities cause the maxima and
minima in the plots. The density gradient decreases with increasing
radius. In the RLOF case the accretion radius increases with distance,
while in the BHL case the behavior of the accretion radius is more
complicated, reaching a maximum after periastron,
depending on the value of $\beta$ in our model and the eccentricity.
The dependance of the density gradient on orbital phase depends on the
value of $\beta$ and $e$ (because the connection between time and
distance depends on $e$), and the accretion radius depends on $e$ in
the RLOF case, and on $e$ and $\beta$ in the BHL case.}
\label{jacc}
\end{figure}

The angular momentum accreted in each step is
\begin{equation}
dJ_{\rm acc} \simeq - \sum_{R_{A}}\sum_{\varphi} (R_{A} \cos \varphi)^2
\Omega dM_A, \label{ja2}
\end{equation}
and the specific angular momentum of the accreted mass in each step
\begin{equation}
j_{\rm acc}(\theta)= \frac {dJ_{\rm acc}}{dM(\theta)}.
\label{ja3}
\end{equation}
The question is whether this mass is accreted directly onto the
secondary. For that, the specific angular momentum is given in units of
the specific angular momentum of a particle in a circular Keplerian motion on
the equator of the secondary $j_2=(G M_2 R_2)^{1/2}$.
The angular momentum accretion rate as function of time is given in
Figure \ref{jacc} for the eight cases.
Summing $dJ_{\rm acc}$ over all steps during the accretion phase gives
the total angular momentum accreted $J_{\rm acc}$, and the specific
angular momentum of the accreted mass is $j_{\rm t-acc}=J_{\rm
acc}/M_{\rm acc}$. The quantity $j_{\rm t-acc}/j_2$ for the total
accreted mass is given in the 6th column of Table \ref{Table:Macc}.

We see from Table \ref{Table:Macc} that the specific angular momentum
of the accreted mass is $j_{\rm t-acc} \simeq j_2$.
Calculating the specific angular momentum of the accreted mass for a
case where the secondary is in a circular orbit with a radius equal to
the periastron distance according to the prescription given by Wang (1981)
gives an average value $j_{A3}/j_2= 2.2$ for the four BHL cases.
This further suggests that indeed $j_{\rm t-acc} \simeq j_2$, and the angular momentum
of the accreted mass cannot be neglected.

By `accretion belt' we refer to mass residing at the surface near the
equator, where centrifugal forces are comparable to, or larger than,
the force due to the thermal pressure gradient, but on average are not
much larger than the gravitational force. Because the accreted mass has
much larger specific angular momentum than the envelope, the envelope
is not deformed to a highly oblate structure, but rather the accreted
mass forms a belt in the equator vicinity. The condition for the
centrifugal force on the accreted mass near the secondary stellar
equator to be important is $j_{\rm t-acc} \gtrsim j_{\rm{2}}$
Our results (Table \ref{Table:Macc}) show that the accreted material
has enough angular momentum to form an accretion belt around the
secondary close to periastron passage.

It is interesting to note that using the primary's wind model and mass
loss rate from Hillier et al. (2001;2006), as described in section
\ref{sec:accmass}, we obtained $j_{\rm t-acc} \simeq 1.6 j_2$.
We conclude that an accretion belt is likely to be formed in this model as well.

\section{RADIATIVE BRAKING}
\label{sec:radiative}

We show here that although radiative braking might have some influence
on the flow, it cannot prevent accretion. For WR+OB (or LBV+OB) binary
systems in which the momentum of the WR wind substantially exceeds that
of the OB star, the WR wind might penetrate into the acceleration zone
of the OB-stellar wind. In such cases it becomes relevant to consider
the role of the OB-stellar radiation in providing the momentum balance
against the WR wind (Gayley et al. 1997). Gayley et al. (1997) found
that a ``sudden radiative braking'' of the WR wind can occur under some
conditions, e.g, Tuthill et al. (2007) found that in WR 104 radiative
braking significantly alters the simple colliding wind geometry. We
follow Tuthill et al. (2007) and Gayley et al. (1997), and apply their
results to \astrobj{$\eta$ Car}.

We take the secondary wind velocity to obey a $\beta$-law with
$\beta=1$. The primary wind velocity will be taken as the terminal
velocity ($500 \km \s^{-1}$) in this analysis (Tuthill et al. 2007).
The primary/secondary wind momentum ratio in \astrobj{$\eta$ Car} is
\begin{equation}
P_{12} = \frac{{\dot M}_{1} v_{1}}{{\dot M}_{2} v_{2}} = 5 .
\label{P12}
\end{equation}
We define $d(\theta)=r(\theta)/R_2$ as the orbital separation at phase
$\theta$ scaled with the secondary radius.
The parameter $\tilde{\eta}$ used by Gayley et al. (1997) is
\begin{equation}
\tilde{\eta}=\frac{4}{3}  \left ( \frac{L_{2}}{L_{1}} \right )^{2}
\left ( \frac{V_{esc,1}}{V_{1}} \right )^{2}  \frac{R_{1}}{R_{2}} = 0.49 , 
\label{etatilde}
\end{equation}
where $V_{esc,1}$ is the primary's escape velocity.

Gayley et al. (1997) also defined the scaled momentum ratio
\begin{equation}
{\hat{P}} = \frac{P_{12}}{P_{rb}} = \frac{5}{1.15} = 4.35,
\end{equation}
and the scaled separation
\begin{equation}
{\hat {d}} = \frac{d(\theta)}{d_{rb}} = \frac{d(\theta)}{2.79} ,
\end{equation}
where $d_{rb}$ is obtained from the cubic-like equation
$d_{rb}=1+\left(\frac{d_{rb}}{\tilde{\eta}}\right)^{1/3}$ and
$P_{rb}={4\beta^\beta d_{rb}^2}/{(2+\beta)^{(2+\beta)}}$.

A system with radiative braking is defined by two conditions:
(1) ${\hat {d}} > 1$, which ensures that the momentum of the secondary
radiation is sufficient to keep the primary wind from impacting the
secondary's surface.
(2) ${\hat {P}} > {\hat {d}}^2$, which implies that there
can be no normal ram pressure balance between the winds, and thus that
radiative braking must stop the primary wind.

The results of our calculation shows that the first condition  ($\hat
{d}>1$) is satisfied along the entire orbital period, but the second
one is not satisfied at all. Therefore there will be no radiative
braking in the common model for \astrobj{$\eta$ Car}.

Under the assumptions of Gayley et al. (1997) the result ${\hat {d}} >
1$ implies that the primary wind will not reach the secondary surface.
However, they neglect gravity because they assume a low density wind. The
very dense primary's wind in \astrobj{$\eta$ Car}, much denser than in
the other systems studied by Gayley et al. (1997), implies that the
radiative braking is less important, and as we now show, it cannot
prevent accretion.

Let us assume that all the accreting star's luminosity is absorbed by the
shocked wind, and the secondary be the accreting mass star. The
momentum discharge (scalar addition of momentum) to the wind is then
$F_{\rm rad}=L_2/c$, which is the radial force exerted by the
secondary's radiation. Let the shocked primary wind be located at a
distance $D_{g2}$ from the secondary. The gravitational force exerted
by the secondary is $F_G=GM_2 \Delta m/D^2_{g2}$, where near periastron
$D_{g2} \simeq 0.5 \AU$. The amount of mass in the shocked region
$\Delta m$ is estimated as follows. A fraction $\zeta$ of the primary
wind is shocked near the secondary. In the case of \astrobj{$\eta$ Car}
$\zeta \sim 0.2$. The post-shock velocity is $\sim v_{\rm wind1}/4$;
near the stagnation point it is much lower, becoming larger at larger
distances. Therefore, the time when the mass leaves the interaction region
is $\tau_f \simeq D_{g2}/(v_{\rm wind1}/4)$, and $\Delta m \simeq \zeta
\dot M_1 \tau_f$. The gravitational force is therefore
\begin{equation}
F_G \simeq \frac{4 G M_2 \zeta \dot M_1}{v_{\rm wind1} D_{g2}}.
\label{fg1}
\end{equation}
Scaling parameters we find the ratio of gravitational to radiation
force to be
\begin{equation}
\begin{split}
\frac{F_G}{F_{\rm rad}} \simeq 1.2 \left( \frac{M_2} {30 M_\odot}
\right) \left( \frac{L_2} {10^6 L_\odot} \right)^{-1} \left(\frac{\zeta} {0.2} \right)
\\
\left( \frac{\dot M_1} {3\times 10^{-4} M_\odot \yr^{-1}} \right)
\left( \frac{D_{g2}} {0.5 \AU} \right)^{-1} \left( \frac{v_{\rm wind1}}
{500 \km \s^{-1}} \right)^{-1}
\label{fg2}
\end{split}
\end{equation}
The ratio is even larger because only a fraction of the secondary
radiation hits the shocked region toward the primary star. Near the
stagnation region, where the post-shock velocity is very low, this
ratio will be much larger. In addition, thermal (and other)
instabilities will form dense blobs, for whom this ratio is larger
even. In a previous paper (Soker 2005) it was already shown that the
radiation pressure cannot stop the accretion of blobs. What makes
\astrobj{$\eta$ Car} different from most other WR-OB systems is the
very dense wind, i.e., a slow high mass loss rate wind, that makes
gravity important. We summarize by noting that although $\hat{d}>1$
along the entire orbit, gravity will ensure accretion close to
periastron passage.

We note here that during the Great Eruption of the 19th century the
mass loss rate was more than 1000 times higher, with $\dot M_1 \sim 1
M_\odot \yr^{-1}$. In that case $F_g > F_{\rm rad}$ up to a distance of
$D_{g2} \sim 100 \AU$, larger than the apastron distance. Indeed,
during the Great Eruption accretion was likely to occur along the
entire orbit (Soker 2005). In that long accretion process, the
accretion disk had time to be established. This lead to the formation
of two jets that shaped the homunculus (Soker 2001).

\section{THE RECOVERY FROM ACCRETION}
\label{sec:dissipation}

Damineli (2008b) composed the `collapse' phase from a rapid fading
phase lasting 3 weeks, a minimum of two months, and a recovery phase of
6 months. In the accretion model the accretion phase lasts $\sim
10$~weeks ending at the end of the X-ray minim phase. Damineli et al.
(2008b) write that there is a general collapse of the wind-wind
collision shock during the collapse phase, but they give no detail to
the physical processes involved. The collapse of the collision region
of the two winds was mentioned before (Soker 2005; Akashi et al. 2006).
As was argued in these papers, and as we do here, this collapse will
inevitably lead to accretion of mass by the secondary star, lasting
$\sim 10$~weeks. More than that, the accreted mass will take the
secondary out of its pre-event structure. The recovery of the secondary
from the accretion event will take several months, accounting for the
half a year recovery phase.

We consider two recovering processes: (1) Removing the accretion belt
around the secondary by its wind. (2) Dense opaque secondary wind that
causes a reduction in the ionizing flux. We discuss these two processes
in the following subsections.

\subsection{Removing the belt}
\label{sec:remove}

Main sequence stars accreting high-entropy mass can expand (Hjellming
\& Taam 1991). For the expansion to be of any significance, say more
than one percent of the radius, the accreted mass must be large if the
results of Hjellming \& Taam (1991) are typical. This might have been
the case during the Great Eruption, when the secondary of
\astrobj{$\eta$ Car} might have accreted several solar masses (Soker
2001, 2007b, 2008). However, during one periastron passage the accreted
mass is tiny, $\sim 10^{-7}$ times the secondary mass. The departure of
the secondary from its equilibrium is only because of the high specific
angular momentum of the accreted mass (section \ref{sec:angular}). It
forms a belt around the secondary because the viscous time is long
(equation \ref{Tvisc}). Because of the long viscosity time and the high
mass loss rate, this belt is destroyed mainly by mass loss rather than
accretion on the secondary.

The accreted mass during a periastron passage is $M_{\rm acc}$, and the
belt covers a fraction $\delta$ of the secondary stellar surface.
For example, if this belt extends from the equator to latitudes $\pm 30^\circ$,
then $\delta=0.5$. We assume that the mass loss rate per
unit solid angle from the belt is as that from the secondary. The belt
will be blown away during a time
\begin{equation}
\begin{split}
t_{\rm belt}=\frac{M_{\rm acc}}{\delta \dot M_2} \simeq  5 \left(
\frac{M_{\rm acc}}{2 \times 10^{-6} M_\odot} \right) \\
\left(\frac{\dot M_2}{10^{-5} M_\odot \yr^{-1}} \right)^{-1}
\left(\frac{\delta}{0.5} \right)^{-1} {\rm month}.
\label{belt1}
\end{split}
\end{equation}
If the mass loss process starts $\sim 60$~days after the event starts,
then the recovery ends $\sim 7$~months after the event starts.
There are large uncertainties, and we consider the value of 5 months close enough to
the half a year recovery phase mentioned by Damineli et al. (2008b).
Namely, the recovery phase is composed of both the time requires for the secondary to get
out of the dense primary wind, for the secondary wind to rebuild itself, and, the longest
process, the ejection of the matter accreted with high specific angular momentum
during the accretion phase.

We considered in here and in previous sections the accretion phase and
the recovery phase. But as has been shown in previous papers on the
accretion model, the behavior of the system depends on the orbital
separation as well, even when the influence of the accretion does not
exist any more. The accretion model does not argue that changes occur
only during the accretion phase and the recovery phase. The accretion
is another process occurring near periastron, but definitely not the
sole process. The variation in the orbital separation was considered as
well in the study of the X-ray (Akashi et al. 2006), Radio (Kashi \&
Soker 2007a), and visible band (Soker 2007a), where good matches to
observations along the entire orbit were obtained. Therefore, the claim
made by Damineli et al. (2008a) against the accretion model is
unjustified.

\subsection{A dense secondary wind}
\label{sec:densewind}

Here we consider the possibility that the secondary wind is very dense,
such that the effective photosphere at radius $R_{wp}$ is determined by
the radius where the wind optical depth becomes $\tau=2/3$
\begin{equation}
\tau_w=\int_{R_{wp}}^\infty \rho(r) \kappa dr=\frac{2}{3}. \label{tau1}
\end{equation}
The wind density is $\dot M_{2,w} (4 \pi r^2 v_w)^{-1}$, where $\dot M_{2,w}$
is the mass loss rate which is assumed to be higher than the regular
mass loss rate $\dot M_2$. We take the velocity profile similar to that
in equation (\ref{v1}), $v_w(r) \simeq v_{wt} [1-(R_\ast/r)]^{\beta}$,
where $R_\ast$ is the stellar radius and $v_{wt}$ the terminal wind
speed, which we can take to be $v_{wt}=v_2$. Equation (\ref{tau1}) can
be integrated analytically to yield
\begin{equation}
\frac{2}{3}=\frac{\kappa \dot M_{2,w} }{4 \pi  (1-\beta) v_{wt} R_\ast}
\left[ 1-\left(1-\frac{R_\ast}{R_{wp}} \right)^{1-\beta} \right].
\label{tau2}
\end{equation}

We scale numerical values as follows. We take the hot star effective
temperature to be $T_{\rm eff} \simeq 40,000 \K$. For the effective
radius we require that the effective temperature of the wind's
photosphere be $\sim 30,000 \K$. At this temperature the photon flux
above the ionization energy of helium is $20 \%$ of that at $T_{\rm
eff} \simeq 40,000 \K$. The radius is $R_{wp} \simeq (4/3)^2 R_\ast
\simeq 1.75 R_\ast$. For a luminosity of $9 \times 10^5 L_\odot$ the
secondary radius is ${R_{wp}}=35 R_\odot$. For the opacity we take
$\kappa= 0.34 \cm^2 \g^{-1}$. Substituting numerical values in equation
(\ref{tau2}) gives for the wind's mass loss rate requires to form the
wind's photosphere at ${R_{wp}}$
\begin{equation}
\dot M_{2,w} = 2.9 \times 10^{-4} \left( \frac{R_{wp}}{35 R_\odot}
\right) \left( \frac{v_wt}{3000 \km \s^{-1}} \right) \Gamma M_\odot
\yr^{-1}, \label{tau3}
\end{equation}
where we define
\begin{equation}
\Gamma = \left[ 1-\left(1-\frac{R_\ast}{R_{wp}} \right)^{1-\beta}
\right]^{-1} \frac{R_\ast}{R_{wp}} (1-\beta) . \label{tau4}
\end{equation}

\begin{figure}[!ht]
   \centering
   \parbox{\textwidth}{
      \hspace{+0.0cm}\includegraphics[width=0.49\textwidth]{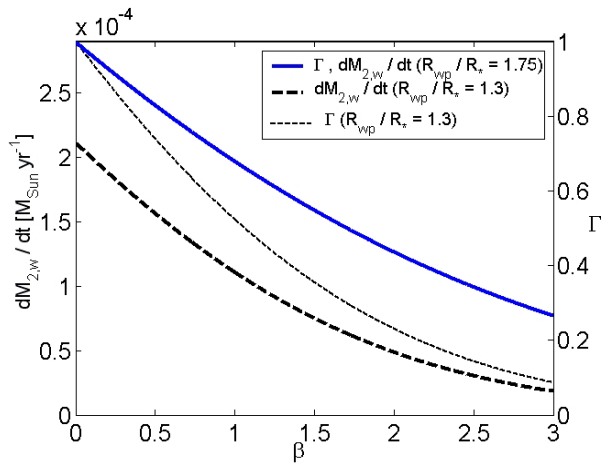}}
\caption{\footnotesize The dependence of $\dot M_{2,w}$ and $\Gamma$ on
      the value of $\beta$, according to equations \ref{tau3} and \ref{tau4},
      respectively.  Solid blue line: $\Gamma$ (scales on the right)
      and $\dot M_{2,w}$ (scales on the left) for $R_{wp}/R_\ast=1.75$.
      Dashed black thin: $\Gamma$ for $R_{wp}/R_\ast=1.3$
      (scales on the right). Dashed black thick: $\dot M_{2,w}$ for $R_{wp}/R_\ast=1.3$.
      Note that $\Gamma$ and $\dot M_{2,w}$ for the case $R_{wp}/R_\ast=1.3$ could not be
      represented by the same line together with $\dot M_{2,w}$ for $R_{wp}/R_\ast=1.75$
      because the value of $R_{wp}$ is different in the two cases.}
 \label{MdotwGammabeta}
\end{figure}

In Figure \ref{MdotwGammabeta} we plot the dependence of $\dot M_{2,w}$ and $\Gamma$
on the value of $\beta$, for two ratios of $R_{wp}/R_\ast=1.75$.

We consider the following scenario. To reduce the helium ionizing flux
by factor of two, the temperature should decrease by a factor of $\sim
1.14$ in the relevant range, e.g., from $\sim 40,000$ to $\sim 35,000$.
This implies a wind photospheric radius of $R_{wp}=1.3 R_\ast=26
R_\odot$. {}From equation (\ref{tau3}) we find that the mass loss rate
(assuming for now that the wind is spherically symmetric) for $\beta=2$
should be $4.9 \times 10^{-5} M_\odot \yr^{-1}$. This higher mass loss
rate can be concentrated near the equatorial plane, where the accretion
belt is. As stated earlier, our approach to the accretion process is
conservative, in the sense that we took a `pessimistic' numerical
values. It is quite possible that the total mass accreted is $M_{\rm
acc}\simeq 10^{-5} M_\odot$, rather than our typical value of $M_{\rm
acc}\simeq 2 \times 10^{-6} M_\odot$. Considering the higher mass loss
rate assumed in the present scenario, the same time is required to
expel the belt, about half a year. To summarize this subsection, in the
scenario discussed in this subsection it is the opaque wind that causes
the reduction in the ionizing flux, rather than the larger radius
formed by the accretion belt.

\subsection{Connection to the recovery from the 19th century eruptions}
\label{sec:nineteen}

Humphreys et al. (2008) discuss the absence of detectable He I lines in
\astrobj{$\eta$ Car} until 1943. The He I intensity was below a quarter of
its present value. We will propose a scenario to account for this
observation in the frame of the binary model.

There were two eruptions in the 19th century: the Great Eruption of
1837-1858 and the Lesser eruption of 1888-1895 (Davidson \& Humphreys
1997). In the binary model for the shaping of the Homunculus the
secondary of \astrobj{$\eta$ Car} accreted several solar masses during
the Great Eruption (Soker 2001, 2007b, 2008). Most likely some
accretion occurred also during the Lesser Eruption. The accreted mass
had high specific angular momentum, and the secondary emerged from the
Great Eruption with a fast rotating outer envelope. In the binary
model, therefore, both the primary and the secondary are recovering
from the two 19th century eruptions (Kashi \& Soker 2008). It is quite
plausible that for $\sim 50 \yr$ after the Lesser Eruption the
secondary had high mass loss rate, about an order of magnitude larger
than its present value. Rapidly rotating OB stars develop a relatively
thick convective region, in particular in the equatorial region (Maeder
et al. 2008). The high mass loss rate for a time period of $\sim 50
\yr$ after the Lesser Eruption might be connected to this convective
region and magnetic activity of the fast rotating secondary. During
that time angular momentum was lost from the outer envelope by the wind
and by transporting it to deeper regions in the star. As a result of
that the outer envelope region spun down. This might explain the
decline in mass loss rate.

{}From equation (\ref{tau3}) we find here that if the secondary hot
star has a mass loss rate higher than its present value by a factor of
$10-15$ and $5-10$, for a secondary wind profile having $\beta=2$
and $\beta=3$, respectively, then the secondary ionizing flux of He I is
less than $20 \%$ of its present value. We suggest that until 1941 the
secondary star had indeed a mass loss rate higher by about an order of
magnitude than its present values. This might explain the weak He I lines until 1941,
whether the lines are formed in the secondary's wind (Kashi \& Soker 2007b),
or in other regions in the binary system (Humphreys et al. 2008).

We conclude that the binary model can explain the absence of He I lines
until $\sim 50 \yr$ after the Lesser Eruption, when previously
suggested ingredients of the model are considered. These are the
accretion of mass with high specific angular momentum by the secondary
during the 19th eruptions, mainly in the Great Eruption, and the
consequently implication that not only the primary, but the secondary
as well is recovering from the eruptions.
It is impossible to predict the time after accretion ceases when an OB star will
start to produce ionizing radiation.
This is because we need to know how the mass loss rate depends on rotation
and magnetic activity, how angular momentum evolves in the envelope, and what is
the magnetic activity of rapidly rotating OB stars.
These three processes are poorly known, and might be highly stochastic.

\section{ALTERNATIVE BINARY MODELS}
\label{sec:alternative}
\subsection{Dense secondary-wind model}
\label{sec:dense}

As stated in section \ref{sec:intro} and as used by us here, the common
model is that the more massive and luminous star, the primary, is the
cooler B star that blows the slow dense wind, while the secondary less
massive star is the hotter component, usually an O star, that blows the
fast wind. This model fits many observations, e.g., Hillier et al.
(2001; 2006) could fit many properties of the primary star in that
model, and Verner et al. (2005) explored the properties of the
secondary. However, some observations are problematic. Consider the He
I visible lines. The spectral fit of Hillier et al. (2001; see Hillier
et al. 2006 for improvements in the model) predicts much too strong
absorption in these lines. In addition, the Doppler shift of thess lines
along the orbit can be accounted for if they are formed in the
atmosphere of the low mass secondary (Kashi \& Soker 2007b), rather
than in the atmosphere of the more massive star.

We consider now the speculative alternative possibility that the
less massive star, the secondary, is the cool star that blows the
dense slow wind, while the more massive star, the primary, blown the
fast wind. This model is studied by Kashi et al. (2008), where the
motivations to consider such a model are given. In addition, we
consider this model because our present study aims at more general
cases of massive binary systems where one of the stars blows a dense
wind. This star can be the more massive or the lighter star in the
binary.

The commonly assumed case where the massive star blows the
slow-dense wind is termed the \emph{dense primary-wind version},
while the case where the secondary less massive star blows the
slow-dense wind is termed the \emph{dense secondary-wind version}.

In the speculative dense secondary-wind version of the model the
massive star still was the one that erupted in the 19th century. After
the eruption it contracted, and now slowly recovers from the eruption.
The secondary accreted $\sim 6-10 M_\odot$ (Soker 2007b, 2008), part of
which was ejected in jets and shaped the Homunculus. The accreted mass
had high specific angular momentum and caused the secondary to become a
fast rotator. This fast rotation might account for the secondary high
mass loss rate in this speculative scenario.

Because of the secondary's fast rotation and the very high wind opacity
the effective secondary's radius must be larger than the $\sim 20
R_\odot$ for an undisturbed late main sequence star. This resembles
Hillier et al. (2001; 2006) model for the primary, where the primary
radius is $60 R_\odot$, but the effective radius we use in modelling the
wind is larger. There is another consideration that suggests an
extended wind acceleration zone. The ratio of the wind momentum
discharge $\dot M v_2$ to that of the radiation $L_2/c$ in the
alternative model is quite large $\sim 8$. This implies an extended
acceleration zone even when there is no rotation. We therefore
conservatively extend the secondary radius to $R_2=30 R_\odot$ in the
dense secondary-wind version.

We apply the alternative model to the accretion process. During most of
the orbital period the basic physics of the collision between the two
winds is indifferent to which star blows which wind.
However, near periastron both the gravitational attraction, which depends on
the stellar mass (Soker 2005), and the acceleration
zone of the slow wind, which depends on stellar mass, radius, and luminosity,
becomes significant in determining the accretion process, and whether it occurs at all.
In the alternative scenario it is the more massive star that accretes during periastron passage
according to the accretion model. As its mass is $\sim 4$ times that of
the secondary, its accretion radius is larger, and practically reaches
the secondary near periastron. On the other hand the high luminosity
implies that the primary radiation pressure will act against accretion.
Substituting the typical values of the common-model of $M_1=120
M_\odot$ and $L_1=4.5 \times 10^6 L_\odot$ in equation (\ref{fg2})
shows that radiative braking is still not an important process.
Therefore, as is the case in the common model,
radiation cannot prevent accretion in the alternative scenario.
As explained in Soker (2005), instabilities in the winds
collision region will form blobs that will be accreted by the primary.

We brought this discussion to emphasize our claim that no matter which
star blows the slow dense wind, its companion is very likely to accrete
from this wind for several weeks during periastron passage. In general,
when the wind is dense such that $F_G>F{\rm rad}$ (equation \ref{fg2}) and the
stars come close enough to each other, accretion is very likely to occur.

\subsection{The big-masses model}
\label{sec:bigmass}

The masses of the stars in the common model are fixed by the assumption
that each star is at its Eddington luminosity. However, fitting to
evolution of main sequence stars will give much higher masses. As
discussed in Kashi et al. (2008) it is quite possible that a model that
better fits the observation is one where the two stars have masses in
the ranges $M_1 \simeq 150-180 M_\odot$ and $M_2 \simeq 60-70 M_\odot$.
Stellar evolutionary models can fit the luminosity of each star, $L_1
\simeq 4.5 \times 10^6 L_\odot$ and $L_2 \simeq 9 \times 10^5 L_\odot$,
but the present masses are determined also by the net mass lost by each
star during the evolution, and by mass accretion by the secondary during
the Great Eruption.
Assuming that the strong binary interaction
increases the mass loss rate, we assume a substantial mass loss rate
and take $(M_1,M_2)= (160,60) M_\odot$. We term this model the
\emph{big-masses model}. We assume at this stage that the radii of the
two stars are the same as in the small-mass model.

As with the small-masses model, there are two versions to the big-masses
model: The commonly assumed case where the massive star blows the
slow-dense wind, the \emph{dense primary-wind version}, and the case
where the secondary less massive star blows the slow-dense wind,
\emph{the dense secondary-wind version}.

In Table \ref{Table:Macc2} we present the results for the big-masses
model, dense primary-wind version. For cases 3 and 4, where $e=0.93$
the BHL accretion radius got into the primary, producing inaccurate
results. Because the masses are larger while the winds properties and
luminosities are the same as those in the small-masses model, a higher
accretion rate is expected. This is indeed the case, as can be seen by
comparing Tables \ref{Table:Macc} and \ref{Table:Macc2}. The average
accreted mass for the eight big-masses model cases we checked is
$M_{\rm acc} \simeq 5 \times10^{-6} M_\odot$, a larger value than for
the small-masses model (the common model). Recalling the discussion in
section \ref{sec:mass}, this implies that about half a year is required
for the secondary to blow away the equatorial belt. This period is
compatible with being the recovery phase; the phase during which some
spectral lines recover (Damineli et al. 2008b).
\begin{table}[!ht]
\begin{tabular}{||c||c|c|c|c|c||}
\hline \hline
Case&Accretion&$e$&$\beta$&$M_{\rm{acc}}$&$\underline{j_{\rm t-acc}}$\\
 &Mode& & &$(10^{-6} M_{\odot})$&${j_2}$\\
\hline \hline

1&BHL&$0.9$&$1$&$1.63$&$1.25$\\

2&BHL&$0.9$&$3$&$10.0$&$3.93$\\

3&BHL&$0.93$&$1$&$\sim2.57$&$\sim2.34$\\

4&BHL&$0.93$&$3$&$\sim18.5$&$\sim4.62$\\

5&RL&$0.9$&$1$&$1.20$&$0.77$\\

6&RL&$0.9$&$3$&$2.32$&$0.84$\\

7&RL&$0.93$&$1$&$1.39$&$0.86$\\

8&RL&$0.93$&$3$&$3.84$&$0.82$\\

\hline \hline
\end{tabular}
\caption{\footnotesize \rm{The accreted mass and specific angular
momentum for the big-masses model, dense primary-wind version. The
columns are parallel to Table \ref{Table:Macc}. $j_{\rm t-acc}$ is
given in units of the specific angular momentum of a circular Keplerian
orbit on the secondary equator $j_2$, which is $\sqrt{2}$ times larger
for the big-masses model than for the small-masses model. For cases 3
and 4 the BHL accretion radius got into the primary, producing
inaccurate results.}} \label{Table:Macc2}
\end{table}

\begin{figure}[!ht]
\resizebox{0.39\textwidth}{!}{\includegraphics{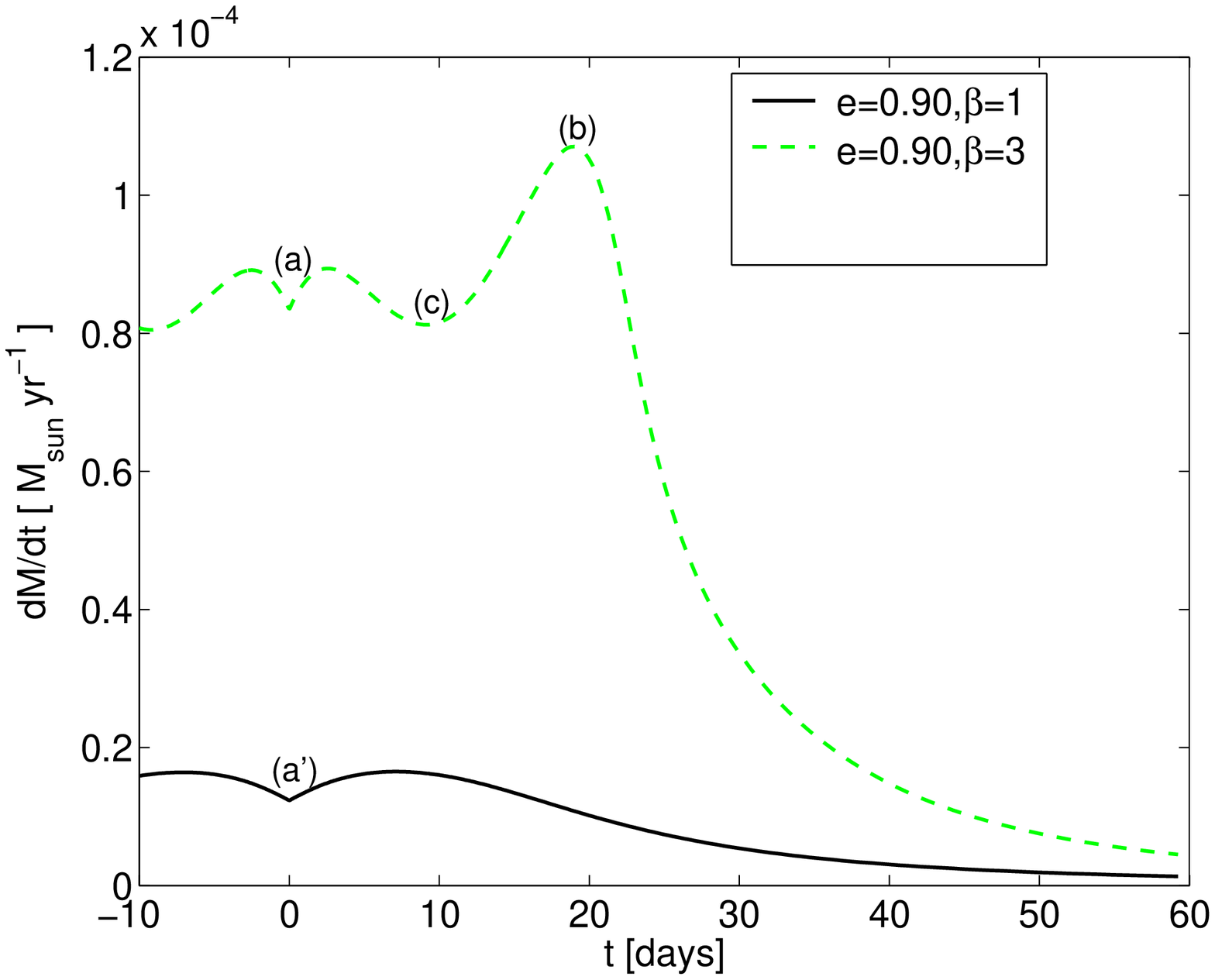}}
\resizebox{0.39\textwidth}{!}{\includegraphics{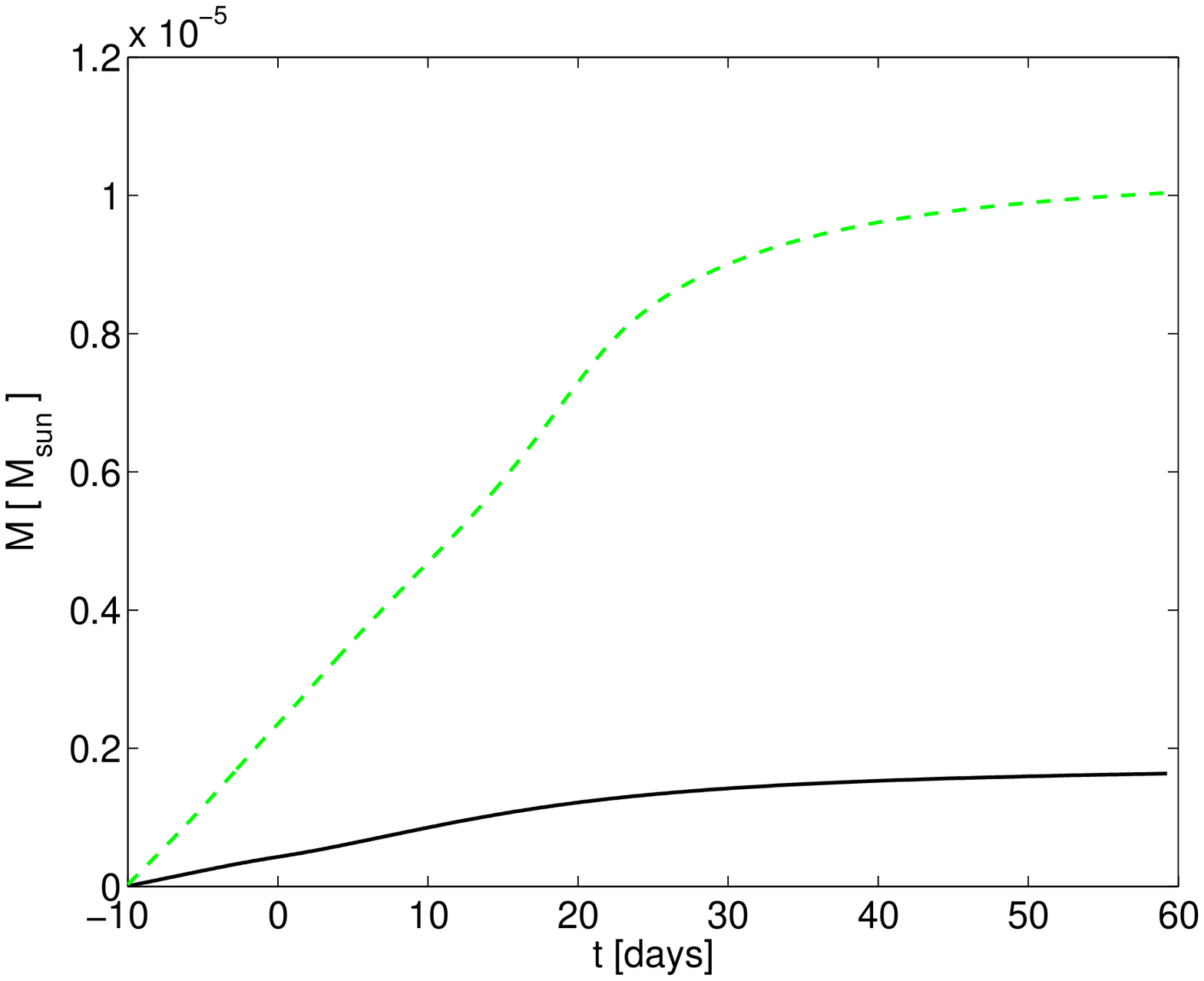}}
\caption{\footnotesize The accretion rate (first panel) and accreted
mass (second panel) as a function of the time near periastron passage,
for the big-masses model, obtained using Bondi-Hoyle-Lyttleton
accretion radius. Only the two cases with $e=0.9$ are plotted, because
for $e=0.93$ the BHL accretion radius is too larger, and the RLOF case
must be used very close to periastron. The maxima in the plots of the
first panel are explained as follows. Near periastron the density is
high, hence the broad maximum around $t=0$. Very close to periastron
the relative velocity between the wind and accreting star is large, and
hence accretion radius small. This explains the deep at $t=0$ (points
$a$ and $a^\prime$) As the two stars recede each other the relative
velocity between the wind and accreting star reach a minimum, and
therefore the accretion radius reaches a maximum (see equation
\ref{bondi}). This explains the maximum in point $b$, on the dashed
line ($e=0.9$, $\beta=3$).} \label{BHL_b}
\end{figure}

\begin{figure}[!ht]
\resizebox{0.39\textwidth}{!}{\includegraphics{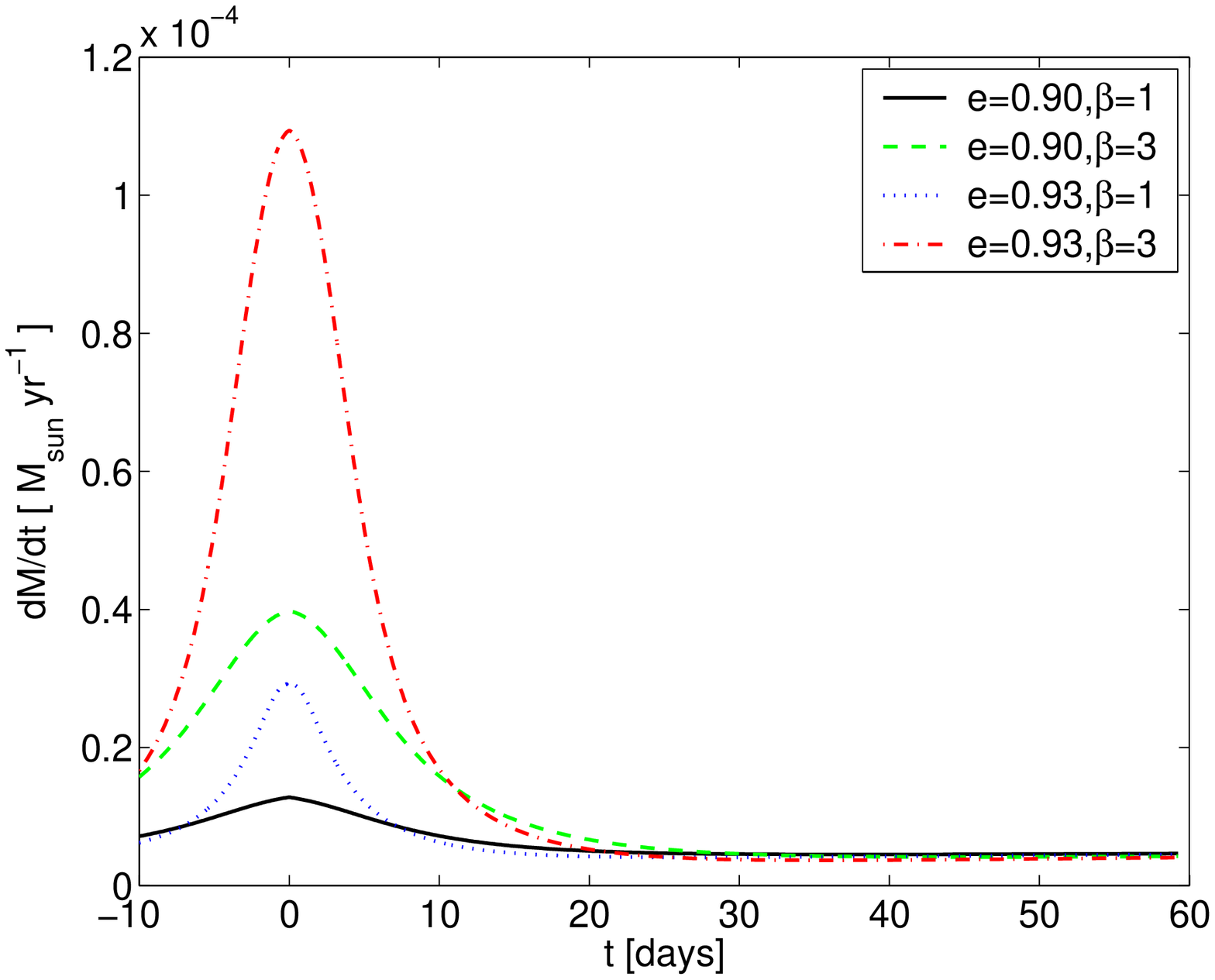}}
\resizebox{0.39\textwidth}{!}{\includegraphics{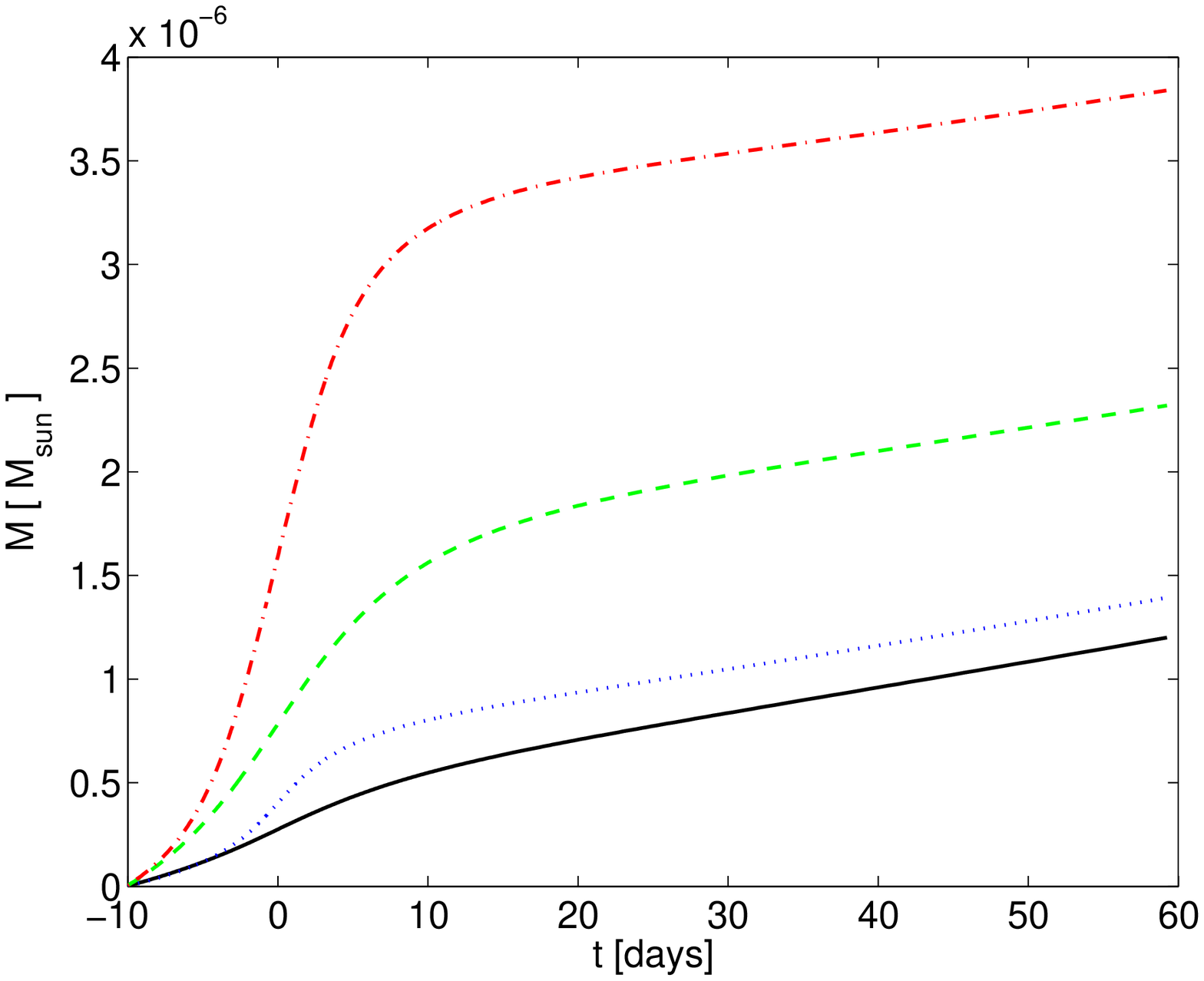}}
\caption{\footnotesize The accretion rate (first panel) and accreted
mass (second panel) as a function of the time near periastron passage,
for the big-masses model, obtained using secondary Roche lobe as
accretion radius (last four rows in Table \ref{Table:Macc2}). Near
periastron the density is high, hence the maximum around $t=0$ in the
mass accretion rate (first panel).
At late times of the accretion phase, $t \ga 10~$day,
the BHL accretion applies.}
\label{RL_b}
\end{figure}

\begin{figure}[!ht]
\resizebox{0.39\textwidth}{!}{\includegraphics{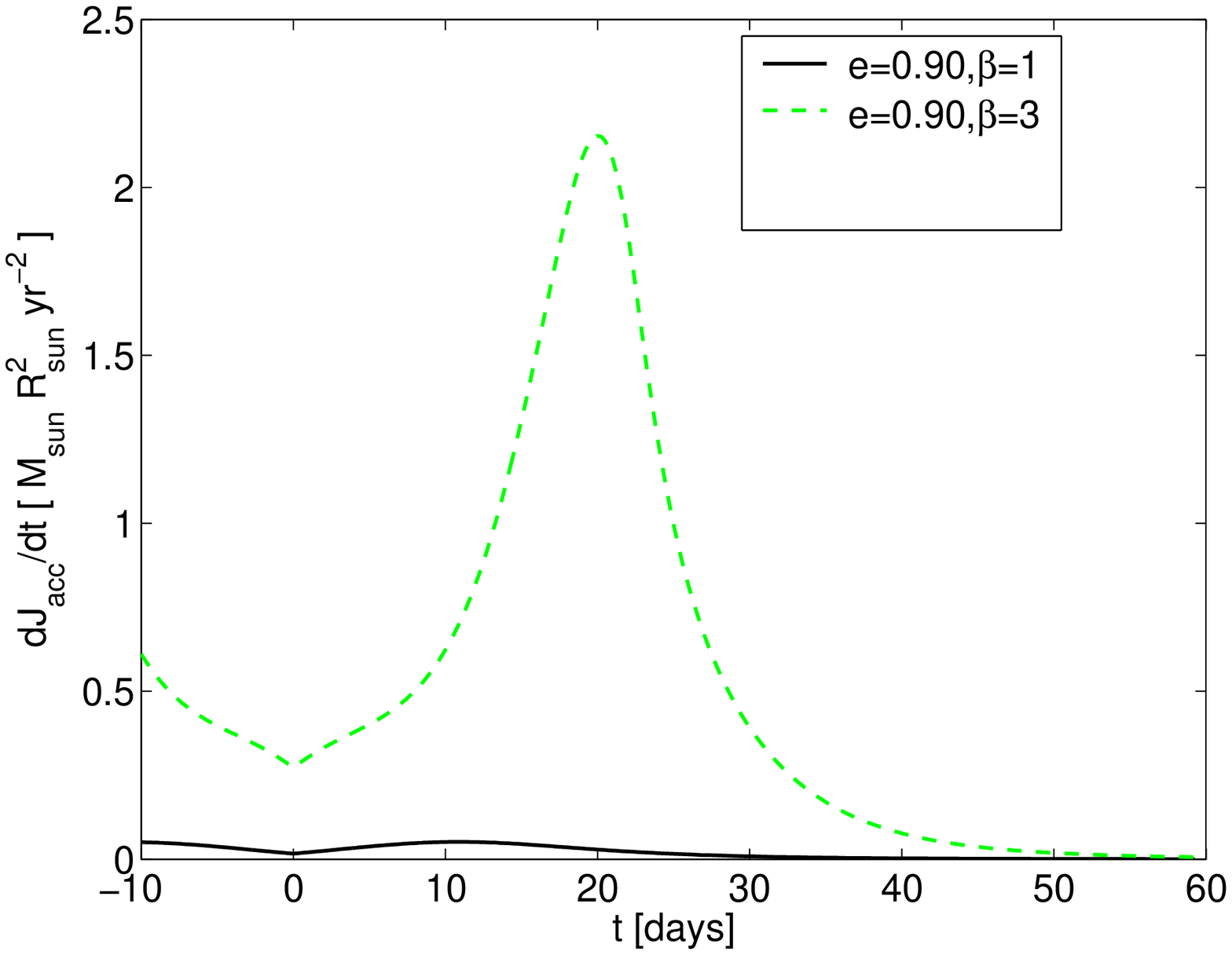}}
\resizebox{0.39\textwidth}{!}{\includegraphics{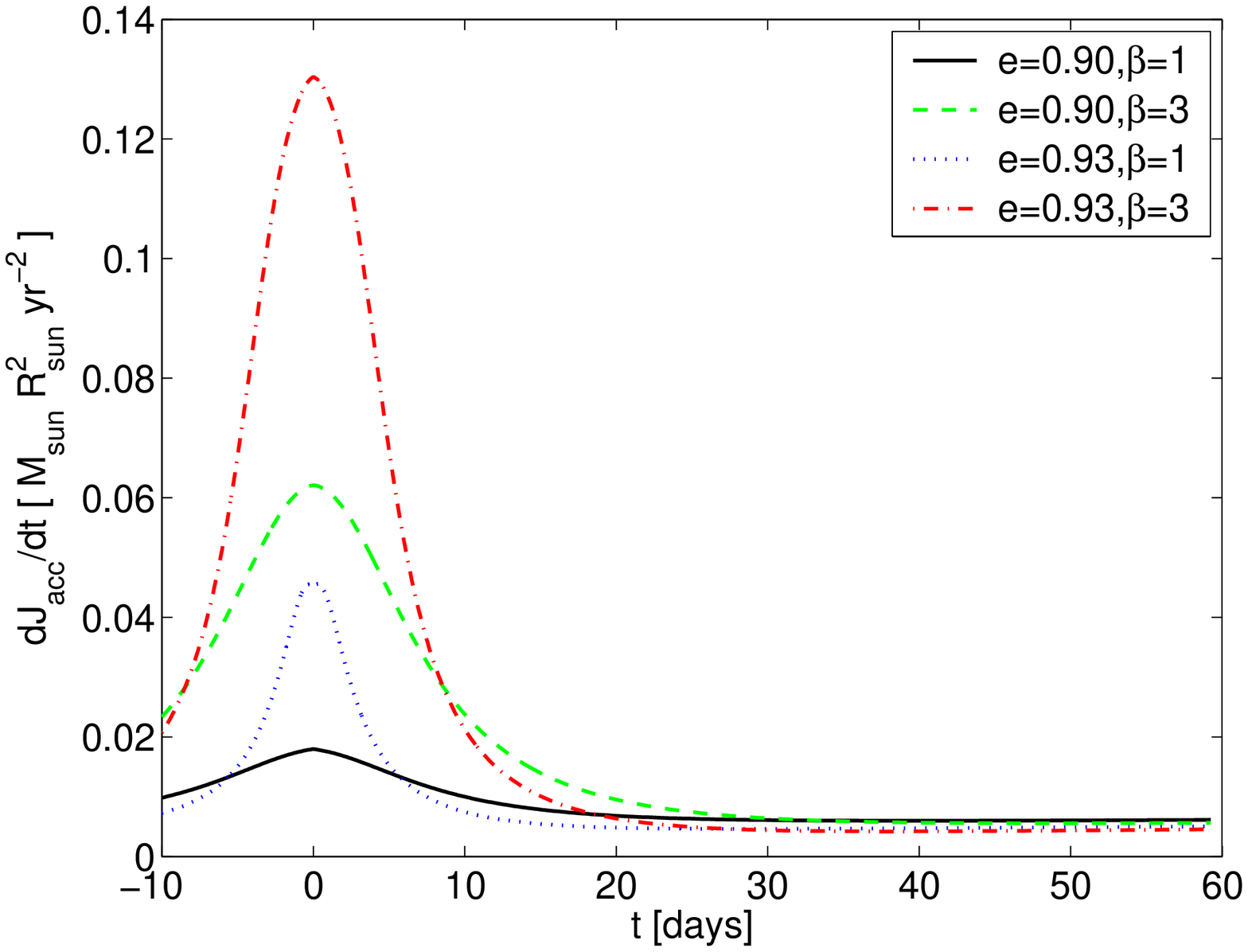}}
\caption{\footnotesize The angular momentum accretion rate as function
of time for the big-masses model. First panel- BHL (only the two cases
with $e=0.9$ are plotted, because for $e=0.93$ the BHL accretion radius
is too larger, and the RLOF case must be used very close to
periastron), second panel- RLOF. The meaning of the different lines is
like in Figure \ref{BHL_b}. The minima and maxima are explained by the
variations in the accretion radius and in the density gradient, both of
which determine the angular momentum of the accreted gas (see caption
to Fig. \ref{jacc}).} \label{jacc_b}
\end{figure}

The behavior of the accretion for the big-masses model is presented in
Figures \ref{BHL_b}-\ref{jacc_b}. These figures are parallel in their
meaning to Figures \ref{BHL}-\ref{jacc}.

The value of $j_{\rm {t-acc}}$ is higher in the big-masses case.
Because $j_2$ is larger as well, the value of
$j_{\rm {t-acc}}/j_2$ might be a little smaller.
$j_2$ is larger by a factor of $\sqrt{(60 M_\odot/30 M_\odot)}=\sqrt{2}$.
We recall that $j_{\rm t-acc}$ is the specific angular momentum of the
total accreted mass (after 10 weeks), and $j_2$ is the specific angular
momentum of a particle performing a circular Keplerian orbit on the
secondary equator.
In the big-masses case the accretion radius is larger, for either BHL or RL cases.
Consequently, the inner part gets very deep into the primary wind,
where the density is very high.
The result is a higher specific angular momentum of the accreted mass.
The value of $j_{\rm {t-acc}} \sim j_2 $ implies that
a belt will form around the accreting star.
The relevance of radiative braking, discussed in section
\ref{sec:radiative}, was checked for the big-masses model. As for the
common model, there will be no radiative braking in that case because
${\hat {P}} < {\hat {d}}^2$ for the entire period. More than that, the
stronger gravity, because of the larger masses, for the same luminosity
implies that radiation cannot prevent accretion.

We also made calculations for the dense secondary-wind version of the
big-masses model. As with the small-masses case, in the dense
secondary-wind version of the big-masses model the radiation cannot
prevent accretion.

Overall, in the big-masses model more mass is accreted during each
periastron passage. Namely, in the big-masses model, whether it is the
primary or secondary that accretes the mass, the accretion model for
the spectroscopic event is more robust.

\section{DISCUSSION AND SUMMARY}
\label{sec:summary}

\subsection{The accretion assumption}

Before summarizing our results we comment on the usage of the accretion
model. In this paper we consider the accretion model where for $\sim
10$~weeks near periastron passage the fast wind-blowing star is
accreting mass from the slow dense wind. The accretion phase starts
$\sim 0-20$~days before periastron passage, and lasts for a total of
$\sim 70$~days.
What about the possibility that the secondary wind is influenced by
the wind interaction near periastron, but no accretion occur,
a non-accretion model.

In the non-accretion model the shocked region of the fast wind is pushed
very close to its origin near the surface of the star, i.e., into its acceleration
zone. Consequently, the fast wind is shocked before it reaches its
terminal speed of $\sim 3000 \km \s^{-1}$, and hot gas at $kT > 5 \keV$
almost does not form, but rather a large mass of cooler gas at $\sim 1
\keV$ is formed. This is compatible with the much lower emission
measure (EM; defined as density square times volume) of the hot gas and
much larger EM of the cooler gas at $kT \sim 1 \keV$ during the X-ray
minimum, relative to their values at most other times (Hamaguchi et al.
2007).

The main problem with the non-accretion model is the asymmetrical
behavior of the X-ray minimum around the event. We assume that
periastron occurs near the beginning of the event as defined by
Damineli et al. (2008a). The X-ray decline under this assumption
starts only several days before periastron passage, and lasts for a
much longer time, $\sim 60$~day after periastron passage.
If it was only for a balance between the ram pressures of the two winds
the X-ray minimum after the event would have been \emph{shorter}. This
is because as the two stars recede each other after periastron passage
the relative speed of the primary wind and the secondary is much lower
than the relative speed before the event (Akashi et al. 2006).
Consequently, the colliding region of the two winds is further away
from the surface of the star blowing the fast wind, and the fast wind
recovers within several days after periastron passage. This
consideration suggests that there is a qualitative change in the
behavior of the secondary wind.

Another asymmetry around the event is in the column density, $N_H$, of
the absorbing material toward the X-ray emitting gas. Even after the EM
of the hot gas increases back to its normal value, at phase 1.042
(Hamaguchi et al. 2006), the value of $N_H$ is very high, and much
larger than its value just before the event, both toward the hot and
toward the cooler X-ray emitting gas (Hamaguchi et al. 2006). Because
the absorbing gas originates in the slow dense wind, the asymmetry in
$N_H$ suggests that there is a qualitative chance in the flow structure
of the dense wind around the winds interaction region.

The accretion model accounts for this behavior.
During the accretion phase
the fast wind is suppressed, but not completely, as accretion is not
spherically symmetric, but rather it is expected to be mainly from the
equatorial plane. Therefore, the fast wind continues in the polar
directions. The large suppression of the secondary wind by the
accretion process implies that the ram pressure of the primary wind
must substantially decrease before the weaker secondary wind can push
back the winds colliding region and rebuild itself. This explains the
late recovery of the hot X-ray emitting gas.

In the accretion flow the slow dense wind forms a wide and dense
accretion column behind the accreting star (Soker 2005),
and a dense material in all directions around the accreting star
(possibly excluding the polar directions). This material is much denser
than in the colliding wind scenario. This possibility is further
explored in Kashi et al. (2008).

In principle, the non-accretion model can account for the asymmetric
behavior of the event if an enhanced mass loss rate from the primary is
induced by the tidal interaction (Pittard \& Corcoran 2002). Such a
process, if occurs, is considered in the frame of the accretion model
as well (Soker 2005). Actually, such an enhanced mass loss rate will
make a denser slow wind, and therefore will increase the accretion
rate. Therefore, it seems that accretion is inevitable if the slow wind
becomes denser even.

We further note the following.
At phase 1.028 during the X-ray minimum, Hamaguchi et al. (2007) find
the largest EM of the cooler gas. The temperature is the lowest out of
their measurements, with a temperature of $kT=0.5 \keV$. The soft
component at $kT=0.5 \keV$ can be formed from the post shock material
of a wind with a speed of $\sim 650 \km\s^{-1}$. For the fast wind with
its terminal speed of $3000 \km \s^{-1}$ the stagnation point of the
colliding wind at phase 1.028 is at a distance of $D_{g2} \simeq 1.7 \AU$
from the center of the secondary star (Akashi et al. 2006).
If the mass loss rate of the fast wind is
unchanged, then the stagnation point for the undeveloped fast wind with
a speed of $\sim 650 \km\s^{-1}$ will be at a distance of
$D_{g2-u} \simeq 0.8 \AU \simeq 9 R_2$.
The fast wind is shocked closer to the secondary. Still, even if the wind
is shocked at a distance of $\sim 0.3 D_{g2-u} \simeq 3 R_2$, the
fast wind will reach a speed larger than 0.25 times its terminal speed.
As mentioned above, it is hard to see how the fast wind will not rebuild
itself at phase 1.02 in a case where its mass loss rate is not suppressed.

We instead suggest that the source of the high EM gas with $kT=0.5 \keV$
is a collimated polar flow blown by the accretion disk, amd/or the shocked slow wind.

\subsection{Results}

Under the assumption that accretion starts 10 days before periastron
and lasts for 60 days after periastron passage, we calculated the
accreted mass and average specific angular momentum for several cases.
The different models discussed in the paper, their properties and
calculated results are presented in Table \ref{Table:sum}. We emphasize
that the values given for $M_{\rm{acc}}$ and $j_{\rm t-acc}/j_{\rm accretor}$
($j_{\rm accretor}$ stands for $j_1$ or $j_2$)
only crudely represent of the values obtained for the eight (only four in
the model of Hillier et al. 2001;2006) subcases of each model and version
as discussed in the previous sections.
The average values don't have a meaning by themselves.

\begin{sidewaystable}
\begin{tabular}{||l|l|l|c|c|c|c|c|c|c||}
\hline \hline
Model&Dense&Accretor&$M_1$&$M_2$&$R_1$&$R_2$&$\beta$&$M_{\rm{acc}}$&$j_{\rm t-acc}$\\
 &Wind& &$(M_{\odot})$&$(M_{\odot})$&$(R_{\odot})$&$(R_{\odot})$& &$(10^{-6} M_{\odot})$&$\overline{j_{\rm{accretor}}}$\\
\hline \hline
Common,small-masses&Primary&Secondary&$120$&$30$&$180$&$20$&$1,3$&$1.66$&$0.89$\\
Alternative,small-masses&Secondary&Primary&$120$&$30$&$52$&$30$&$1,3$&$4.06$&$1.60$\\
Common,big-masses&Primary&Secondary&$160$&$60$&$180$&$20$&$1,3$&$5.19$&$1.93$\\
Alternative,big-masses&Secondary&Primary&$160$&$60$&$52$&$30$&$1,3$&$5.56$&$2.03$\\
Hillier et al. (2001;2006)&Primary&Secondary&$120$&$30$&$99.4$&$20$&$4.08$&$1.07$&$0.57$\\
\hline \hline
\end{tabular}
\caption{\footnotesize \rm{A summary of the five different models
discussed in the paper. The input parameters, listed from left to right:
Name of model, the star blowing the slow dense wind, the star that accretes the mass near periastron,
the masses and radii of the two stars. The primary is always the more massive stars.
$\beta$ is the parameter of the wind velocity profile as given in equation (\ref{v1}).
The results of the calculations are the accreted mass $M_{\rm{acc}}$,
and the average specific angular momentum $j_{\rm t-acc}$. In the last column we mark
whether a belt (or an accretion disk) is formed.
  The values given for the total mass accreted in the assumed 10 weeks
long accretion phase $M_{\rm{acc}}$ and the average specific angular
momentum $j_{\rm t-acc}$ are only crude averages. The average specific
angular momentum $j_{\rm t-acc}$ is given in units of the specific
angular momentum of a particle performing a Keplerian circular orbit on
the equator of the accretor $j_{\rm accretort}$. Radiative braking was
not important for neither of the models.}} \label{Table:sum}
\end{sidewaystable}

Our reference model is the commonly assumed model, the common model,
where the primary blows the slow dense wind, and the masses of the two
stars are $(M_1, M_2) \simeq (120,30) M_\odot$. We term the common
model the dense primary-wind version of the small-masses model.

Our main results can be summarized as follows.
\begin{enumerate}
\item For eccentricity of $e \sim 0.9-0.95$ the accreting star is most likely to accrete
 $\sim 10^{-6} M_\odot$ during each periastron passage.
\item The specific angular momentum of the accreted mass is larger than, or
about equal to, the value that is required to perform a Keplerian
motion around the accreting star. Therefore, in most cases an accretion
disk will be formed (beside may be in the dense secondary-wind version
of the small-masses model; see Table \ref{Table:sum}).
However, the viscosity time scale is long, and the
disk will not reach an equilibrium. In any case, most of the accreted
mass will be accreted from regions close to the equatorial plane, and a
belt of gas will be formed near the accreting star equator. The
effective temperature on the surface of the belt region is lower, and
UV radiation softer.
\item The viscosity time scale is too long (section \ref{sec:timescales}),
and it is not efficient in transporting angular momentum and in dissipating
the thick disk (or belt). We assume that it is the wind from the accreting star, which
rebuilds itself after the accretion phase, that expels the mass in the
belt. The ejection of the belt by the wind requires several months. We
suggest to identify this time period with the half a year recovery
phase of some spectral lines.
\end{enumerate}

Our accretion model, where mass is accreted for several weeks near
periastron passages, is not commonly accepted by the community. Our
goal here is to show that even under conservative assumptions and
approximations, mass accretion is very likely to occur.
Namely, even when we neglect the tidal force of the secondary on the primary,
and when we consider only the tangential motion very close to
periastron, we cannot avoid the conclusion that the secondary
gravitationally accretes mass for several weeks.

\subsection{Implications for the next periastron passage}

The most clear prediction of the model is the presence of an accretion
disk, or a belt if the disk is close to the equator of the accreting
star and is not in equilibrium (because the viscosity time is longer
than the duration of the accretion phase). Accretion disks have clear
signatures, like broad and/or double peaked emission lines. However,
the same dense wind that is accreted obscures the relevant regions.
Therefore, we must examine indirect indications.

(1) {\it The duration of the accretion phase.} The accretion process
strongly depends on the eccentricity of the orbit, the mass of the
accreting star, and the winds properties. The masses and eccentricity
have negligible changes from orbit to orbit. However, stellar winds are
known to be stochastic. Therefore, it is possible that the slow dense
wind will be stronger or weaker in the next event. This will result in
a change in the duration of the  the X-ray minimum. In the last two
cycles the two X-ray minima last for the same time, with only small
variations at the beginning and end of the minimum. This indicates that
the winds did not change much between these two cycles. If the winds
properties will not change much by January 2009, then the X-ray minimum
will last for $\sim 10$ weeks again. However, if significant changes
will occur in the winds, mainly in the slow dense wind, then we expect
the duration of the x-ray minimum to be shorter or longer, for a weaker
or a stronger slow dense wind, respectively. Because of the fast
variation in orbital separation close to periastron, the change in the
X-ray minimum duration will be of only several days.

(2) {\it Polar outflow.} The accretion from the equatorial region
leaves the polar directions clear for the secondary weaker wind, or for
a polar outflow blown by the accretion disk. Again, the visible and UV
bands are obscured. However, hard X-ray emission might be detected. In
particular, such a polar outflow might be strong when the fast wind
builds itself. We encourage a search for Doppler shifts in X-ray
emission and absorption lines as the systems exits the X-ray minimum.
Most emission from the red-shifted polar outflow will be absorbed, and
we expect for a blue shifted emission during that time. We note that
hints for a polar outflow exist from X-ray emission lines several weeks
before the event (Behar et al. 2007). Several weeks before the event is
before the accretion phases has started. Only accretion of dense blobs
is possible.

(3) {\it Polar illumination. } The accretion disk, or belt, will
survive for weeks to months after the accretion ends. During that time
the accreting star will illuminate and ionize the polar directions
much more than in the equatorial plane. This effect has already been
observed by Stahl et al. (2005) and Weis et al. (2005) who found that the
spectroscopic event is more pronounced in the equatorial and mid
latitudes directions than in the polar directions. We expect this
effect to be strong during the accretion phase, coincide with the X-ray
minimum, and to continue for weeks after.

(4) The X-ray observations presented by Hamaguchi et al. (2007) are not
frequent enough to present the evolution of the X-ray emission of the
cooler component (below 5 Mev). We hope frequent observations will be
made during the X-ray minimum of the next cycle (January to March
2009). As mentioned above, we suggest that the origin of part of this gas with
high EM and $kT \simeq 0.5 \keV$ is a disk outflow. We therefore
predict that the high EM of this gas will be built during the high
accretion rate time, phase 0.995-1.01, when the disk builds itself.

We thank the referee for useful comments.
This research was supported in part by the Asher Fund for Space Research at the
Technion.


\end{document}